\documentclass[reqno,12pt]{amsart}

\usepackage{amsmath,amssymb,amsthm,amsfonts,amsxtra}
\usepackage{fullpage}

\numberwithin{equation}{section} 

\newtheorem{thm}{Theorem}{\bf}{\em}
{\bf}{\em}
\newtheorem{prop}{Proposition}{\bf}{\em}
\newtheorem{lem}{Lemma}{\bf}{\em}
\newtheorem{rem}{Remark}{\bf}{\em}
{\bf}{\em}

\def\ov{\bar}

\def\pe{\perp}
\def\pa{\parallel}

\def\mb{\mathbf}
\def\mr{\mathrm}
\def\mk{\mathfrak}
\def\bs{\boldsymbol}

\def\beq{\begin{equation}}
\def\eeq{\end{equation}}

\def\bpm{\begin{pmatrix}}
\def\epm{\end{pmatrix}}

\def\bel{\begin{lem}}
\def\eel{\end{lem}}
\def\ber{\begin{rem}}
\def\eer{\end{rem}}
\def\bep{\begin{prop}}
\def\eep{\end{prop}}

\def\Rnum{\mathbb{R}}
\def\Cnum{\mathbb{C}}

\DeclareMathOperator{\re}{Re}
\DeclareMathOperator{\im}{Im}
\DeclareMathOperator{\spn}{span}
\DeclareMathOperator{\tr}{tr}

\DeclareMathOperator{\ad}{ad}
\DeclareMathOperator{\Ad}{Ad}

\def\t{{\rm t}}
\def\cc{{\mathcal C}}
\def\inv{{}^{-1}}
\def\id{{\rm id}}

\def\<{\langle}
\def\>{\rangle}

\def\brack#1{\langle #1\rangle}

\def\const{{\rm const.}}

\def\gsp{\mk{g}}
\def\msp{\mk{m}}
\def\hsp{\mk{u}(n)}

\def\asp{\mk{a}}

\def\mpasp{\msp_{\pa}}
\def\mpesp{\msp_{\pe}}
\def\hpasp{\hsp_{\pa}}
\def\hpesp{\hsp_{\pe}}

\def\i{\mr{i}}
\def\j{\mr{j}}

\def\e{\mr{e}}
\def\norme{\chi}
\def\efac{\tfrac{1}{\sqrt{\norme}}}
\def\efacsq{\norme^{-1}}

\def\h{\mr{h}}

\def\wedg#1#2{#1\wedge #2}
\def\wedgbar#1#2{#1\,\bar\wedge\, #2}
\def\symm#1#2{#1 \odot #2}
\def\symmbar#1#2{#1\,\bar\odot\, #2}

\def\conx{\omega}
\def\hook{\rfloor}
\def\JJ{\mb{J}}

\def\ee{\hat{\mb{e}}}

\def\u{\mr{u}}

\def\w{\mr{w}}

\def\uu{{\mb u}}
\def\uubar{\ov{\uu}}
\def\ww{{\mb w}}
\def\wwbar{\ov{\ww}}

\def\hh{{\mb h}}
\def\hhbar{\ov{\hh}}

\def\TTh{\bs{\Theta}}
\def\thone{\Theta_1}
\def\thtwo{\Theta_2}
\def\thtwobar{\ov{\Theta}{}_2}

\def\HH{{\mb H}}
\def\HHbar{\ov{\HH}}

\def\lrep{{\bs(}}
\def\rrep{{\bs)}}

\def\aa{{\mb a}}
\def\bb{{\mb b}}
\def\cc{{\mb c}}
\def\dd{{\mb d}}
\def\aabar{\ov{\aa}}
\def\bbbar{\ov{\bb}}

\def\aaone{{\mb a}_1{}}
\def\aatwo{{\mb a}_2{}}
\def\bbone{{\mb b}_1{}}
\def\bbtwo{{\mb b}_2{}}
\def\ccone{{\mb c}_1{}}
\def\cctwo{{\mb c}_2{}}
\def\ddone{{\mb d}_1{}}
\def\ddtwo{{\mb d}_2{}}
\def\aaonebar{\ov{\mb a}_1{}}
\def\aatwobar{\ov{\mb a}_2{}}
\def\bbonebar{\ov{\mb b}_1{}}
\def\bbtwobar{\ov{\mb b}_2{}}
\def\cconebar{\ov{\mb c}_1{}}

\def\ddtwobar{\ov{\mb d}_2{}}

\def\AA{\mb{A}}
\def\BB{\mb{B}}
\def\CC{\mb{C}}
\def\DD{\mb{D}}
\def\AAbar{\ov{\AA}}
\def\BBbar{\ov{\BB}}
\def\CCbar{\ov{\CC}}

\def\a{\mr{a}}
\def\b{\mr{b}}
\def\c{\mr{c}}
\def\d{\mr{d}}

\def\bone{\mr{b}_1{}}
\def\btwo{\mr{b}_2{}}

\def\btwobar{\ov{\mr b}_2{}}
\def\done{\mr{d}_1{}}
\def\dtwo{\mr{d}_2{}}

\def\dtwobar{\ov{\mr d}_2{}}

\def\Rop{{\mathcal R}}
\def\Ropbar{\ov{\mathcal R}}
\def\Jop{{\mathcal J}}
\def\Hop{{\mathcal H}}
\def\Kop{{\mathcal K}}
\def\Eop{{\mathcal E}}

\def\X{{\rm X}}

\def\map{\gamma}

\def\secref#1{Sec.~\ref{#1}}

\def\Ref#1{Ref.~\cite{#1}}

\def\eg/{e.g.}
\def\ie/{i.e.}

\def\scrpt#1{$\scriptstyle {#1}$}

\begin{document}

\allowdisplaybreaks[3]
\tolerance =99999

\title{Unitarily-invariant integrable systems\\ and geometric curve flows in $SU(n+1)/U(n)$ and $SO(2n)/U(n)$}

\author{
Ahmed Ahmed${}^{1,2}$
\lowercase{and}
Stephen C. Anco$^2$
\lowercase{and}
Esmaeel Asadi$^3$
\\
\\
\lowercase{\scshape{
${}^1$
Department of Mathematics\\
University of South Florida\\
Tampa, FL 33620, USA\\
}}\\
\lowercase{\scshape{
${}^2$
Department of Mathematics and Statistics\\
Brock University\\
St. Catharines, ON L\scrpt2S\scrpt3A\scrpt1, Canada\\
}}
\\
\lowercase{\scshape{
${}^3$
Department of Mathematics\\
Institute for Advance Studies in Basic Science (IASBS)\\
45137--66731, Zanjan, Iran\\
}}
}

\thanks{ahmedmgahmed@hotmail.com, sanco@brocku.ca, esmaeel.asadi@gmail.com}

\begin{abstract}
Bi-Hamiltonian hierarchies of soliton equations are derived from 
geometric non-stretching (inelastic) curve flows in the Hermitian symmetric spaces 
$SU(n+1)/U(n)$ and $SO(2n)/U(n)$. 
The derivation uses Hasimoto variables defined by a moving parallel frame along the curves. 
As main results, new integrable multi-component versions of 
the Sine-Gordon (SG) equation and the modified Korteveg-de Vries (mKdV) equation,
as well as a novel nonlocal multi-component version of the nonlinear Schr\"odinger (NLS) equation are obtained, 
along with their bi-Hamiltonian structures and recursion operators. 
These integrable systems are unitarily invariant and correspond to geometric curve flows
given by a non-stretching wave map and a mKdV analog of a non-stretching Schr\"odinger map in the case of the SG and mKdV systems, 
and a generalization of the vortex filament bi-normal equation 
in the case of the NLS systems. 
\end{abstract}

\maketitle

%\tableofcontents
\date{\today,\version}

\section{Introduction and summary}

The nonlinear Schr\"odinger (NLS) equation $u_t = \i(u_{xx} + |u|^2u)$ 
is one of the most prominent examples of an integrable system. 
Its possesses an infinite hierarchy of symmetries 
and a corresponding infinite hierarchy of conservation laws, 
which are generated by a recursion operator. 
It also possesses two different but compatible Hamiltonian structures
and corresponding Poisson brackets. 
As well, it possesses a Lax pair from which an inverse scattering transform arises 
that can be used for solving the initial-value problem. 

All of these integrability aspects of the NLS equation have a remarkable 
geometrical origin when non-stretching (inelastic) curve flows are considered in Euclidean space \cite{NakSegWad,DolSan,MarSanWan,AncMyr}. 
A non-stretching curve flow is an equation $\map_t = a \hat T + b\hat N +c \hat B$
formulated in a Frenet frame $(\hat T,\hat N,\hat B)$ along a curve $\map$ 
with an arclength-parameterization, so that $\map_x=\hat T$ is the unit tangent vector. 
The normal coefficients $b,c$ in the flow equation are functions of 
the curvature $\kappa$ and the torsion $\tau$ of the curve, 
given by the Frenet equations 
$\hat T_x = \kappa \hat N$, $\hat N_x = -\kappa \hat T + \tau \hat B$, $\hat B_x = -\tau N_x$, 
while the tangential coefficient $a$ is determined by $a_x=\kappa b$
due to the non-stretching property of the curve. 
When $b=a=0$ and $c=\kappa$, the curve $\map$ undergoes a bi-normal flow 
$\map_t = \kappa \hat B$. 
This flow equation physically describes the motion of a vortex filament in incompressible fluids \cite{Has}. 
The induced flow on $(\kappa,\tau)$ turns out to be equivalent to the NLS equation 
for the Hasimoto variable $u=\kappa\exp(\i\int\tau\,dx)$ \cite{Has}. 
Moreover, the Lax pair and bi-Hamiltonian operators for the NLS equation 
turn out to be encoded in a simple way in the structure equations of a moving frame formulation of the curve flow \cite{MarSanWan,AncMyr}, 
where the Hasimoto transformation from $(\kappa,\tau)$ to $u$ 
corresponds geometrically to a gauge transformation from a Frenet frame to a parallel frame 
given by rotating the vectors $(\hat N,\hat B)$ in the normal plane by 
an angle $\theta(x) = -\int\tau\, dx$ along the curve \cite{Bis}. 
Unlike a Frenet frame, 
a parallel frame has a rigid gauge freedom consisting of a constant rotation $\phi$ 
applied to the vectors $(\hat N,\hat B)$ in the normal plane. 
Under this rigid gauge transformation, 
$u$ transforms to $e^{\i\phi}u$ by a constant phase rotation,
and so $u$ is not an invariant of the curve like $(\kappa,\tau)$ 
but instead has the geometrical meaning of a $U(1)$-covariant of the curve
\cite{AncMyr}. 

Similarly, all of the symmetries of the NLS equation themselves 
correspond to geometrical curve flows in Euclidean space, 
and in particular the first higher symmetry in the NLS hierarchy is given by 
$\map_t = \kappa\tau\hat B + \kappa_x \hat N + \tfrac{1}{2}\hat T$
with the Hasimoto variable $u$ satisfying 
the complex modified Korteveg-de Vries (mKdV) equation  
$u_t = u_{xxx} + \tfrac{3}{2}|u|^2u_x$. 
This flow equation physically describes axial motion of a vortex filament \cite{FukMiy}. 
It is also an integrable system, sharing the same integrability properties
as the NLS equation. 

A broad generalization of parallel frames and Hasimoto variables 
has been obtained in work over the past few decades on non-stretching curve flows
in more general geometric spaces, 
starting with constant-curvature Riemannian manifolds 
\cite{DolSan,LanPer1998,SanWan,Wan}
$S^n$ (spheres) and $H^n$ (hyperbolic spaces), 
and continuing with various other Riemannian geometries 
\cite{Anc2006a,Anc2006b,Anc2007,AsaSan,AncAsa2009,AncAsa2012,AncAsaDog}
given by symmetric spaces $M=G/H$ (also called reductive Klein geometries). 
As a culmination of this line of work, 
for general Riemannian symmetric spaces $M=G/H$, 
a complete theory of parallel frames, Hasimoto variables, 
and integrable systems of mKdV type as well as Sine-Gordon (SG) type 
arising from geometrical non-stretching curve flows 
was presented in \Ref{Anc2008} by the second author. 
In particular, 
a pair of compatible Hamiltonian operators (Poisson brackets) was shown to be encoded
in the structure equations of a $H$-parallel frame for non-stretching curve flows 
in a general Riemannian symmetric space $M=G/H$. 
This integrability structure provides a recursion operator that yields a hierarchy of mKdV-type  integrable systems, including a SG-type integrable system. 

Riemannian symmetric spaces are curved generalizations of Euclidean space
in which the Euclidean isometry group $SO(3)\ltimes \Rnum^3$ 
is replaced by a (real) simple Lie group $G$,
and the $SO(3)$ gauge group of the frame bundle 
is replaced by a subgroup $H\subset G$ that is invariant under an involutive automorphism of $G$. 
There is a well-known classification of these spaces (see e.g.\cite{Hel}), 
based on the classification of 
real simple Lie groups. 
These respective classifications each have a division into 
classical types,  known as the A, B, C, D series, 
and exceptional types, known as the E, F, G series. 

To-date, mKdV and SG integrable systems,
along with their integrability structure and their corresponding geometrical realizations as non-stretching curve flows,
have been derived in the following classical types:\\
$\bullet$
$SO(n+1)/SO(n)$, BD I \cite{Anc2006b}\\
$\bullet$
$SU(n)/SO(n)$, A I \cite{Anc2006b}\\
$\bullet$
$Sp(n+1)/Sp(1)×Sp(n)$, C II \cite{AsaSan,AncAsa2009,AncAsa2012}\\
$\bullet$
$SU(2n)/Sp(n)$, A II \cite{AncAsa2012}\\
$\bullet$
$Sp(n)/U(n)$, C I \cite{AncAsaDog}

The purpose of the present paper is to consider the two remaining simplest classical types:
\\
$\bullet$ 
$SU(n+1)/U(n)$, A III; and $SO(2n)/U(n)$, D III

In \secref{sec:structeqns}, 
the general theory from \Ref{Anc2008} will be applied to these two Riemannian symmetric spaces. 
As one new development, 
the theory will be extended to derive an NLS-type integrable system 
by exploiting a $U(1)$ subgroup given by the center of the equivalence group of 
the $U(n)$-parallel frame. 
In \secref{sec:SU} and \secref{sec:SO},
the specific features of the resulting integrable NLS systems 
in $SU(n+1)/U(n)$ and $SO(2n)/U(n)$ will be worked out,
as well as the integrable mKdV systems and SG systems, 
along with their integrability structure. 
In the case of $SU(n+1)/U(n)$, 
these integrable systems involve a real scalar variable and a complex vector variable,
whereas in the case of $SO(2n)/U(n)$, 
they involve a real scalar variable and a pair of complex vector variables. 
In \secref{sec:curveflows}, 
the corresponding geometrical non-stretching curve flows will be derived 
for each of these integrable systems. 
Most interestingly, the integrable NLS systems are found to correspond to 
a generalized bi-normal curve flow in $SU(n+1)/U(n)$ and $SO(2n)/U(n)$. 
Finally, some concluding remarks will be made in \secref{sec:remarks}.  

We mention that all of the integrable systems derived here 
also possess a Lax pair that comes directly from the frame structure equations
in a $U(n)$-parallel frame, since these equations have the form of a zero-curvature matrix equation in the Lie algebra of the isometry groups $SU(n+1)$ and $SO(2n)$. 
The details of this will be given in a subsequent paper \cite{AncAsa.preprint} 
in which we will derive 
the Lax pair associated to the structure equations of a $H$-parallel frame 
in a general Riemannian symmetric space $M=G/H$. 

We also point out that the general theory in \Ref{Anc2008}, 
as well as its application in the present paper, 
is different than the work of Fordy and Kullish \cite{ForKul}
in which integrable NLS systems are derived by writing down a linear isospectral problem 
in Hermitian symmetric spaces. 
In particular, those systems essentially utilize the Hermitian structure of the space, 
and as a consequence they have different form and a different number of components 
than the ones we derive. 
Our systems are constructed without using the Hermitian structure of $SU(n+1)/U(n)$ and $SO(2n)/U(n)$, 
since this structure is not part of the equivalence group of the $U(n)$-parallel frame 
that is used in the construction. 
Instead we use an almost complex structure that 
exists in the tangent space of the symmetric space along the curve flow 
and comes from the center of the frame equivalence group. 
Moreover, 
the geometrical realization of the Fordy-Kullish systems involves stretching (elastic) curve flows \cite{LanPer2000}
rather than the non-stretching (inelastic) curve flows that we consider here.

\section{$U(n)$-parallel frames and curve flow equations}
\label{sec:structeqns}

The symmetric spaces $M=G/U(n)$ with $G=SU(n+1),SO(2n)$ 
have a natural Riemannian structure
which comes from a soldering identification between 
the tangent space $T_xM$ at points $x$
and the vector space $\msp=\gsp/\mk{u}(n)$. 
This soldering $T_xM \simeq \msp$ relies on the algebraic properties of 
$\gsp \supset \hsp$ 
as a symmetric Lie algebra. 
In particular, 
\begin{equation}
\gsp=\hsp\oplus\msp, 
\quad
K(\hsp,\msp) =0
\end{equation}
is an orthogonal direct sum decomposition 
relative to the Cartan-Killing form $K$ on $\gsp$, 
with the Lie bracket relations
\begin{equation}\label{brack.rel}
[\hsp,\hsp]\subset \hsp,
\quad
[\hsp,\msp] \subseteq \msp,
\quad 
[\msp,\msp]\subset \hsp 
\end{equation}
induced from the Lie bracket on $\gsp$. 

A simple formulation of the soldering identification is provided by \cite{KobNom} 
a $\msp$-valued linear coframe $e$
and a $\hsp$-valued linear connection $\conx$
whose torsion and curvature
\begin{equation}\label{tor.curv}
\mk{T}:=de+\bs{[}\conx,e\bs{]},
\quad
\mk{R}:=d\conx+\tfrac{1}{2}\bs{[}\conx,\conx\bs{]}
\end{equation}
are $2$-forms with respective values in $\msp$ and $\hsp$,
given by the Cartan structure equations
\begin{equation}\label{cart.stru}
\mk{T}=0,
\quad 
\mk{R}=-\tfrac{1}{2}\bs{[}e,e\bs{]}.
\end{equation}
Here $\bs{[}\cdot,\cdot\bs{]}$ denotes the Lie bracket on $\gsp$
composed with the wedge product on $T_x^*M$.
This structure together with the Cartan-Killing form 
determines a Riemannian metric $g$ 
and a Riemannian connection (\ie/ covariant derivative) $\nabla$ 
on the space $M=G/U(n)$ 
from the following soldering relations:
\begin{gather}
g(X,Y):=-K(e_X,e_Y),
\label{metr.M}\\
e\hook\nabla_X Y:=\partial_X e_Y+[\conx_X,e_Y],
\label{conn.M}
\end{gather}
for all $X,Y$ in $T_x M$,
where $e\hook X=e_X$, $e\hook Y=e_Y\in\msp$.
In addition, 
the $2$-forms $\mk{T}$ and $\mk{R}$ determine 
the torsion tensor $T(X,Y):=\nabla_XY-\nabla_YX-[X,Y]$ 
and the curvature tensor $R(X,Y):=[\nabla_X,\nabla_Y]-\nabla_{[X,Y]}$ 
through the soldering relations
\begin{gather}
e\hook T(X,Y)=\mk{T}\hook(X\wedge Y)=0,
\label{tor.tensor.M}\\
e\hook R(X,Y)Z=[\mk{R}\hook(X\wedge Y),e_Z]=-[[e_X,e_Y],e_Z] . 
\label{curv.tensor.M}
\end{gather}
These expressions \eqref{metr.M}--\eqref{curv.tensor.M}
show that $\nabla$ is metric compatible, $\nabla g=0$, 
and torsion-free, $T=0$,
while its curvature is covariantly constant, $\nabla R=0$. 

This formulation of the Riemannian structure of $M=G/U(n)$ 
has an intrinsic gauge freedom consisting of the transformations
\begin{equation}\label{gauge.transf}
e\longrightarrow \Ad(f^{-1})e,
\quad
\conx\longrightarrow\Ad(f^{-1})\conx+f^{-1}df
\end{equation}
as defined in terms of an arbitrary function $f:M\to U(n) \subset G$.
The gauge transformations \eqref{gauge.transf}
comprise a local ($x$-dependent) representation of
the linear transformation group $\Ad(U(n))$
which defines the gauge group \cite{Sha} of the frame bundle of $M$.
Both the metric tensor $g$ and curvature tensor $R$ on $M$ are gauge invariant.

Geometrically, 
$G$ represents the isometry group of $M$,
while the subgroup $U(n)\subset G$ represents the isotropy subgroup of
the origin $o$ in $M$. 
In terms of the symmetric Lie algebra structure \eqref{brack.rel},
the Lie subalgebra $\hsp$ is identified with the generators of
isometries that leave fixed the origin $o$ in $M$,
and the vector space $\msp$ is identified with the generators of
isometries that carry the origin $o$ to any point $x\neq o$ in $M$.

The symmetric space $M=G/U(n)$ also has a Hermitian structure, 
consisting of a complex structure tensor $J$ that satisfies the properties 
$J^2=-\id$, $\nabla J=0$, and $g(JX,JY)=g(X,Y)$ for all $X,Y$ in $T_x M$. 
In particular, $J$ can be identified with the linear map $\Ad(U(1)_C)$ on $T_xM$ 
where $U(1)_C$ is the center of $U(n)$. 

Let $\map(x)$ be any smooth curve in $M=G/U(n)$.
A moving frame consists of 
a set of orthonormal vectors that span the tangent space
$T_\map M$ at each point $x$ on the curve $\map$.
The Frenet equations of a moving frame yield a connection matrix
consisting of the set of frame components of the covariant $x$-derivative of
each frame vector along the curve \cite{KobNom}.
A moving coframe consists of a set of orthonormal covectors that are dual to
the frame vectors relative to the Riemannian metric $g$.
Such a framing for $\map(x)$ is determined by the Lie-algebra components
of $e$ and $\conx\hook \map_x$
when an orthonormal basis is introduced for $\msp=\gsp/\hsp$
with respect to the Cartan-Killing form,
where the Frenet equations are defined by the frame components of 
the transport equation
\begin{equation}\label{frenet.eq}
\nabla_x e=-\ad(\conx\hook\map_x)e
\end{equation}
along the curve.
In particular,
if $\{\mb{m}_l\}$, $l=1,\ldots,\dim\msp$, 
is any fixed orthonormal basis for $\msp$,
then a frame at each point $x$ along the curve is given by
the set of vectors $X_l:=-\<e^*,\mb{m}_l\>$, $l=1,\ldots,\dim\msp$.
Here $e^*$ is a $\msp$-valued linear frame 
defined as the dual to the linear coframe $e$ 
by the condition that $-\<e^*,e\>=\textrm{id}$
is the identity map on each tangent space $T_xM$
(see \cite{Anc2008,KobNom}).

Now consider any smooth flow $\map(x,t)$ of a curve in $M=G/U(n)$.
We write $X=\map_x$ for the tangent vector
and $Y=\map_t$ for the evolution vector
at each point $x$ along the curve.
The flow is {\em non-stretching} (inelastic) provided that
it preserves the $G$-invariant arclength $ds=|\map_x|dx$,
or equivalently 
\begin{equation}\label{non.stretch}
\nabla_t|\map_x|=0
\end{equation}
in which case we have
$g(\map_x,\map_x)=|\map_x|^2=1$
without loss of generality.
For flows that are transverse to the curve
(such that $X$ and $Y$ are linearly independent),
$\map(x,t)$ will describe a smooth two-dimensional surface in $M$.
The pullback of the torsion and curvature equations \eqref{cart.stru}
to this surface yields
\begin{align}
&
D_xh-D_t\e+[u,h]-[\varpi,\e]=0,
\label{pull.1}\\
&
D_x\varpi-D_tu+[u,\varpi]=-[\e,h],
\label{pull.2}
\end{align}
with
\begin{align}
&
\e:=e\hook X=e\hook \map_x,
\quad
u:=\conx\hook X=\conx\hook \map_x,
\label{pull.tangential}\\
&
h:=e\hook Y=e\hook \map_t,
\quad
\varpi:=\conx\hook Y=\conx\hook \map_t,
\label{pull.flow}
\end{align}
where $D_x,D_t$ denote derivative operators with respect to $x,t$.
In terms of these variables, the curve flow is given by 
\beq\label{curveflow}
\map_t = -K(e^*,h)
\eeq
where the linear frame $e^*$ is determined in terms of $u$ 
from the dual of the Frenet equation \eqref{frenet.eq}, 
\beq\label{dual.frenet.eq}
\nabla_x e^*=-\ad(u)e^* .
\eeq

For any non-stretching curve flow $\map(x,t)$, 
these structure equations \eqref{pull.1}--\eqref{pull.flow}
turn out to encode a pair of compatible Hamiltonian operators,
as shown in Theorem~\ref{thm:HJops} later. 
The encoding looks simplest when we utilize an $U(n)$-parallel framing for $\map(x,t)$ 
as follows. 

A $U(n)$-parallel frame along a curve in $M=G/U(n)$
is a direct algebraic generalization of a parallel moving frame in Euclidean geometry \cite{Bis},
as defined by the properties \cite{Anc2008}:
\begin{enumerate}
\item[(i)]
$\e$ is a constant unit-norm element lying in a Cartan subspace
$\asp\subset \msp$ 
that is contained in the centralizer subspace
$\msp_{\pa}$ of $\e$, \ie/
$D_x\e=D_t\e=0, K(\e,\e)=-1$, and $\ad(\msp_{\pa})\e=0$
where $\msp_{\pa}\oplus\msp_{\perp}=\msp$ 
and $K(\msp_{\pa},\msp_{\perp})=0$.
\item[(ii)]
$u$ lies in the perp space $\hsp_{\perp}$ of the Lie subalgebra
$\hsp_{\pa}\subset \hsp$ of the linear isotropy group
$\Ad(U(n))_{\pa}\subset \Ad(U(n))$ that preserves $\e$, \ie/
$\ad(\hsp_{\pa})\e=0$ and $K(u,\hsp_{\pa})=0$
where
$\hsp_{\pa}\oplus\hsp_{\perp}=\hsp$
and $K(\hsp_{\pa},\hsp_{\perp})=0$.
\end{enumerate}
Existence of such moving frames can be established by applying 
a suitable gauge transformation \eqref{gauge.transf} 
to an arbitrary linear frame
at each point $x$ along the curve \cite{Anc2008}.
The necessary transformation is unique up to a residual gauge freedom 
given by rigid transformations in $\Ad(U(n))_{\pa}\subset \Ad(U(n))$ 
preserving the tangent vector $X$,
where the subgroup $U(n)_\pa\subset U(n)$ is generated by 
the Lie subalgebra $\hsp_\pa\subset \hsp$. 
This residual gauge freedom is called the {\em equivalence group} of the frame. 

The resulting $U$-parallel coframe $e$ provides an isomorphism between
$T_\map M$ and $\msp=\gsp/\hsp$, 
which yields a correspondence between 
the vectors $\{ X_l \}$ in a moving frame for $T_\map M$
and the vectors $\{\mb{m}_l\}$ in a basis for $\msp$.
Under this isomorphism, 
the tangent vector $X=\map_x$ corresponds to the Cartan element $\e=e\hook X\in\asp$. 

\begin{rem}\label{rem:equiv.frames}
The set of inequivalent $U(n)$-parallel frames admitted by a smooth arclength-parameterized curve $\map(x)$ in $M=G/U(n)$ 
can be characterized by the set of orbits of all unit-norm elements $e\hook \map_x =\e$ 
in the Cartan subspace $\asp\subset \msp$ under the action of the subgroup in $U(n)$
that preserves this subspace \cite{Anc2008}. 
The dimension of $\asp$ is equal to the rank of the symmetric space $G/U(n)$,
which is $1$ in the case $G=SU(n+1)$ and $[n/2]$ (integer part) in the case $G=SO(2n)$.
Thus,  up to equivalence, 
a $U(n)$-parallel frame is unique for the symmetric space $M=SU(n+1)/U(n)$, 
but is not non-unique for the symmetric space $M=SO(2n)/U(n)$. 
In both spaces $M$, the equivalence group $U(n)_\pa$ of a $U(n)$-parallel frame 
does not contain the circle group $\exp(J\phi)\subset U(n)$ ($\phi\in\Rnum$)
generated by the complex structure $J$, 
since no vector in $T_xM$ is invariant under this group. 
\end{rem}

The properties and construction of $U(n)$-parallel frames 
rely on the Lie bracket relations 
for the subspaces $\msp_{\pa}$, $\msp_{\pe}$, $\hsp_{\pa}$, $\hsp_{\pe}$
coming from the structure of $\gsp$
as a symmetric Lie algebra \eqref{brack.rel}.
These relations consist of
\begin{align}
&
[\msp_\pa,\msp_\pa] \subseteq \hsp_\pa,
\quad
[\msp_\pa,\hsp_\pa] \subseteq \msp_\pa,
\quad
[\hsp_\pa,\hsp_\pa] \subseteq \hsp_\pa,
\label{liebrac1}\\
&
[\hsp_\pa,\msp_\pe] \subseteq \msp_\pe,
\quad
[\hsp_\pa,\hsp_\pe] \subseteq \hsp_\pe,
\label{liebrac2}\\
&
[\msp_\pa,\msp_\pe] \subseteq \hsp_\pe,
\quad
[\msp_\pa,\hsp_\pe] \subseteq \msp_\pe,
\label{liebrac3}
\end{align}
while the remaining Lie brackets obey the general relations
\begin{equation}
[\msp_\pe,\msp_\pe] \subset \hsp, 
\quad
[\hsp_\pe,\hsp_\pe] \subset \hsp, 
\quad
[\hsp_\pe,\msp_\pe]\subset \msp .
\label{liebrac4}
\end{equation}

\subsection{Non-stretching curve flow equations}

We will consider non-stretching curve flows $\map(x,t)$ in $M=G/U(n)$
having a $U(n)$-parallel framing. 
By projecting the Cartan structure equations \eqref{pull.1}--\eqref{pull.2}
into the subspaces $\msp_\pa$, $\msp_\pe$, $\hsp_\pa$, $\hsp_\pe$, 
we obtain the system 
\begin{align}
& 0 = D_x h_\pa  +[u, h_\pe]_\pa  \in \msp_\pa ,
\label{cartaneq.mpar}
\\
& 0 = D_x h_\pe+ [u, h_\pa]+ [u, h_\pe]_\pa  +[\e,\varpi^\pe] \in \msp_\pe,
\label{cartaneq.mperp}
\\
& 0 = D_x \varpi^\pa  + [u,\varpi^\pe]_\pa  
\in \hsp_\pa , 
\label{cartaneq.hpar}
\\
& 0 = D_x \varpi^\pe - D_t u+[u,\varpi^\pa]+ [u,\varpi^\pe]_\pe +  h^\pe \in \hsp_\pe,
\label{cartaneq.hperp}
\end{align}
where
\begin{equation}
u=u_\pe\in\hsp_\pe,
\quad
h=h_\pa +h_\pe\in\msp_\pa \oplus\msp_\pe,
\quad
\varpi=\varpi^\pa +\varpi^\pe\in\hsp_\pa \oplus\hsp_\pe , 
\end{equation}
and 
\begin{equation}\label{hperp.up}
h^\pe = \ad( \e ) h_\pe \in \hsp_\pe . 
\end{equation}

Through equations \eqref{cartaneq.mpar} and \eqref{cartaneq.hpar}, 
the variables $h_\pa$ and $\varpi^\pa$ can be eliminated, 
while from equations \eqref{cartaneq.hperp} and \eqref{hperp.up}, 
$\varpi^\pe$ can be expressed in terms of $h^\pe$. 
General results in \Ref{Anc2008} show that the resulting equations encode a pair of compatible Hamiltonian operators \cite{Olv,Dor} as follows. 

\begin{thm}\label{thm:HJops}
In a $U(n)$-parallel framing for non-stretching curve flows in $G/U(n)$, 
with $G=SU(n+1),SO(2n)$, 
the frame structure equations take the form 
\begin{equation}\label{flow.eqs.HJ.form}
u_t = \Hop(\varpi^\pe) + h^\pe, 
\quad
\varpi^\pe = \Jop(h^\pe) 
\end{equation}
where $\Hop = \Kop|_{\hpesp}$ is a Hamiltonian operator, 
and $\Jop = \ad(\e)\inv\Kop|_{\mpesp}\ad(\e)\inv$ is a compatible symplectic operator,
given by 
\beq
\Kop = D_x + [u,\cdot]_\pe +[u,D_x\inv[u,\cdot]_\pa]
\eeq
These operators and the flow equation on $u$ 
are invariant under the unitary equivalence group $U(n)_\pa\subset U(n)$ 
of the $U(n)$-parallel frame. 
\end{thm}

%An operator $\Eop$ is Hamiltonian \cite{Olv,Dor} iff the associated bracket 
%$\{\mk{F},\mk{G}\}_\Eop =$
%$\int K(\delta\mk{G}/\delta u,\Eop(\delta\mk{F}/\delta u))\, dx$ 
%is a Poisson bracket (namely, it is skew and obeys the Jacobi identity),
%for all functionals $\mk{F},\mk{G}$ depending on $u$ and $x$ derivatives of $u$. 
%An operator $\Dop$ is symplectic \cite{Dor} iff its inverse $\Dop$ is a Hamiltonian operator. 
%Compatibility of $\Dop$ and $\Eop$ means that every linear combination 
%$c_1\Eop+c_2\Dop\inv$ is a Hamiltonian operator
%and also that $\Eop\Dop$ is a hereditary recursion operator. 

We emphasize that the Hamiltonian operator structure of the flow equation \eqref{flow.eqs.HJ.form} in Theorem~\ref{thm:HJops} 
is universal for all non-stretching curve flows in $M=G/U(n)$.
This has an important application when the flow equation is written in the form 
\begin{equation}\label{flow.eqs.R.form}
u_t = \Rop(h^\pe) + h^\pe, 
\quad
\Rop = \Hop\Jop 
\end{equation}
with $\Rop$ being a hereditary recursion operator. 
By Magri's theorem \cite{Mag}, 
if the flow component $h^\pe$ is chosen to be the generator of a symmetry 
$\X = \h^\pe\hook\partial_u$ of the operators $\Hop$ and $\Jop$, 
then the resulting flow equation on $u$ will be an integrable system
possessing a bi-Hamiltonian structure and a hierarchy of higher symmetries and associated conservation laws. 
Moreover, 
the corresponding curve flow \eqref{curveflow} will be a geometrical flow that will 
possess a similar integrability structure. 

One obvious symmetry of $\Hop$ and $\Jop$ is $x$-translations,
as generated by $\X = u_x\hook\partial_u$.
Hence 
\begin{equation}\label{x.flow}
u_t -u_x= \Rop(u_x)
\end{equation}
is an integrable system. 
It possesses $\Rop$ as a recursion operator,
which generates the hierarchy of higher symmetries
\begin{equation}\label{x.flow.symms}
\X^{(k)} = h^\pe_{(k)}\hook\partial_u,
\quad
h^\pe_{(k)} = \Rop^k(u_x),
\quad
k=0,1,2,\ldots
\end{equation}
starting from the $x$-translation symmetry $\X^{(0)}= u_x\hook\partial_u$. 
Note that the convective term $-u_x$ can be removed by a Galilean transformation. 
Then the resulting system $u_t = \Rop(u_x)$ is of complex mKdV type. 

Another symmetry of $\Hop$ and $\Jop$ consists of unitary transformations on $u$,
which come from the equivalence group $U(n)_\pa\subset U(n)$ of the $U(n)$-parallel frame. 
The center of $U(n)_\pa$ consists of a $U(1)$ subgroup 
whose action on $\msp_\pe$ is isomorphic to that of the circle group 
$\exp(J\phi)\subset U(n)$ ($\phi\in\Rnum$). 
This provides a circle-group symmetry 
which is generated by $\X = (\ad(j)u)\hook\partial_u$ 
where $j\in\mk{u}(1)\subset \hpasp\subset\hsp $ satisfies $\ad(j)^2=-\id$ on $\mpesp$.
Then 
\begin{equation}\label{ad_j.flow}
u_t -\ad(j)u= \Rop(\ad(j)u)
\end{equation}
is an integrable system possessing the hierarchy of higher symmetries
\begin{equation}\label{ad_j.flow.symms}
\X^{(k)} = h^\pe_{(k)}\hook\partial_u,
\quad
h^\pe_{(k)} = \Rop^k(\ad(j)u),
\quad
k=0,1,2,\ldots
\end{equation}
starting from the $U(1)$ symmetry $\X^{(0)}= (\ad(j)u)\hook\partial_u$. 
The term $\ad(j)u$ can be removed by a phase transformation $u\to \exp(t\ad(j))u$,
and this yields a system $u_t = \Rop(\ad(j)u)$ which is of NLS type. 

In addition to the two integrable systems \eqref{x.flow} and \eqref{ad_j.flow}, 
an integrable system of Sine-Gordon type can be obtained by considering 
the kernel of the symplectic operator, 
\beq\label{kerJ.flow}
u_t = h^\pe, 
\quad
\Jop(h^\pe) =0.
\eeq
This system shares the hierarchy of higher symmetries admitted by the mKdV-type integrable system \eqref{x.flow.symms}. 

The specific structure of all three integrable systems \eqref{x.flow}, \eqref{ad_j.flow}, \eqref{kerJ.flow} 
for the two symmetric spaces $G/U(n)$ 
will be discussed in the next two sections.

\section{Integrable systems in $SU(n+1)/U(n)$}
\label{sec:SU}

Since the symmetric space $SU(n+1)/U(n)$ has rank $1$, 
there is a unique choice of a $U(n)$-parallel frame, up to equivalence, 
for non-stretching curve flows $\map(x,t)$ 
(see Remark~\ref{rem:equiv.frames}). 
The equivalence group of this frame is 
$\Ad(U(n-1)) \subset \Ad(U(n))$. 

To construct the $U(n)$-parallel frame, 
we will need the following matrix decomposition of the symmetric Lie algebra 
$\mk{su}(n+1)= \hsp\oplus\msp$:
\begin{equation}\label{su.matr.rep}
\bpm -\tr(\BB)&\aa\\-\aabar^\t&\BB \epm 
\in \mk{su}(n+1),
\quad 
\BB\in\mk{u}(n),
\aa\in\Cnum^n,
\tr(\BB)\in\i\Rnum ,
\end{equation}
where we will write 
\begin{align}
\lrep\BB\rrep :=& \bpm -\tr(\BB)& \mb{0}\\\mb{0}&\BB \epm 
\in\hsp\subset\mk{su}(n+1),
\quad 
\BB\in\mk{u}(n),
\tr(\BB)\in\i\Rnum,
\label{su.hsp.matr.rep}
\\
\lrep\aa\rrep :=& \bpm 0&\aa\\-\aabar^\t & \mb{0} \epm 
\in\msp=\mk{su}(n+1)/\mk{u}(n) \simeq \Cnum^n,
\quad 
\aa\in\Cnum^n .
\label{su.msp.matr.rep}
\end{align}
The Cartan subspace $\asp\subset\msp\simeq \Cnum^n$ is the real span of 
the element $\lrep\ee\rrep$ given by 
$\ee = (1,\mb{0})\in\Cnum^n = \Cnum\oplus\Cnum^{n-1}\simeq\msp$, 
where $\mb{0}\in\Cnum^{n-1}$. 
Using this Cartan element $\lrep\ee\rrep$, 
we obtain a decomposition of 
$\msp=\mpasp\oplus\mpesp$ and $\hsp=\hpasp\oplus\hpesp$ 
into centralizer subspaces and perp subspaces:
\begin{align}
& \aa = (\a_\pa+\i\a_\pe,\aa_\pe) 
\in \Cnum^{n}, 
\quad
\aa_\pe\in\Cnum^{n-1},
\a_\pa,\a_\pe\in\Rnum,
\label{su.msp.perp.pa.matr.rep}
\\
& \BB = \bpm \i\b_\pe -\tfrac{1}{2}\tr(\BB_\pa)& \bb_\pe \\ -\bbbar_\pe^\t & \BB_\pa \epm 
\in \hsp, 
\quad
\BB_\pa\in\mk{u}(n-1),
\bb_\pe\in\Cnum^{n-1},
\b_\pe\in\Rnum,
\tr(\BB_\pa)\in\i\Rnum,
\label{su.hsp.perp.pa.matr.rep}
\end{align}
with the notation 
\begin{align}
& 
\lrep\BB_\pa\rrep := 
\bpm 
-\tfrac{1}{2}\tr(\BB_\pa)& (0, \mb{0}) \\ 
(0,\mb{0})^\t  & 
\bpm -\tfrac{1}{2}\tr(\BB_\pa) & \mb{0} \\ \mb{0}^\t & \BB_\pa \epm
\epm 
\in \hsp_\pa,
\label{su.hsp.pa}
\\
&
\lrep(\i\b_\pe,\bb_\pe)\rrep := 
\bpm 
-\i\b_\pe & (0, \mb{0}) \\ 
(0,\mb{0})^\t  & 
\bpm \i\b_\pe&\bb_\pe\\-\bbbar_\pe^\t&\mb{0}\epm
\epm 
\in \hsp_\pe,
\label{su.hsp.pe}
\\
& \lrep\a_\pa\rrep := \bpm 0& (\a_\pa,\mb{0}) \\ -(\a_\pa,\mb{0})^\t & \mb{0} \epm 
\in \msp_\pa,
\label{su.msp.pa}
\\
& \lrep(\i\a_\pe,\aa_\pe)\rrep:=\bpm 0& (\i\a_\pe,\aa_\pe) \\ (\i\a_\pe,-\aabar_\pe)^\t & \mb{0} \epm 
\in \msp_\pe.
\label{su.msp.pe}
\end{align}
Details of this decomposition and its algebraic properties, 
including the Lie bracket structure, 
are summarized in Appendix~\ref{app:su.liealg}. 

We now write out the frame structure equations \eqref{cartaneq.mpar}--\eqref{cartaneq.hperp} by using these matrix representations \eqref{su.hsp.pa}--\eqref{su.msp.pe}. 

First, the tangential components \eqref{pull.tangential} of 
the linear coframe $e$ and the linear connection $\conx$ 
along the curve are given by the variables
\begin{align}
& \e=\efac \lrep(1,\mb{0})\rrep  \in\msp_\pa\subset\msp,
\label{su.e}
\\
& u= \lrep(\i\u,\uu)\rrep \in\hsp_\pe,
\label{su.u}
\end{align}
where $\u\in\Rnum$ is a real scalar variable, 
and $\uu\in\Cnum^{n-1}$ is a complex vector variable. 
These variables geometrically represent the components of the Cartan matrix of the $U(n)$-parallel frame. 
Here $\norme = 2\sqrt{n+1}$ is a normalization constant determined by the Cartan-Killing metric 
(see Proposition~\ref{prop:su.struct} in Appendix~\ref{app:su.liealg}). 

Next, the flow components \eqref{pull.flow} of 
the linear coframe and the linear connection 
are expressed in terms of the variables
\begin{align}
& h_\pa=\lambda \lrep\h_\pa\rrep \in \msp_\pa,
\label{su.h.par}
\\
& h_\pe=\lrep(\i\h_\pe,\hh_\pe)\rrep  \in\msp_\pe,
\label{su.h.perp}
\\
& \varpi^\pa=\lrep\TTh\rrep   \in\h_\pa, 
\label{su.w.par}
\\
& \varpi^\pe=\lrep(\i\w,\ww)\rrep \in\h_\pe,
\label{su.w.perp}
\end{align}
as well as 
\beq
h^\pe =\lambda\efac \lrep(\i\h^\pe,\hh^\pe)\rrep \in\h_\pe ,
\quad
\h^\pe = -2\lambda^{-1}\h_\pe,
\quad
\hh^\pe = -\lambda^{-1}\hh_\pe
\label{su.h.perp.up}
%\h^\pe= \ad(\e)h_\pe =\efac \lrep(-2\i\h_\pe,-\hh_\pe)\rrep 
\eeq  
from relation \eqref{hperp.up}. 
Here
$\h_\pa$, $\h_\pe$, $\h^\pe$, $\w$ $\in\Rnum$ are real scalar variables; 
$\hh_\pe$, $\hh^\pe$, $\ww$ $\in\Cnum^{n-1}$ are complex vector variables; 
$\TTh$ $\in\mk{u}(n-1)$ is an anti-hermitian matrix variable. 
Then, the frame structure equations \eqref{cartaneq.mpar}--\eqref{cartaneq.hperp}
are given by 
\begin{gather}
D_x\h_\pa =\h^\pe\u +\brack{\hh^\pe,\uu},
\label{su.hpar.eq}
\\
D_x\TTh =\wedgbar{\uu}{\ww},
\label{su.Th.eq}
\\
\efac\lambda^{-1} \w=\tfrac{1}{4} D_x\h^\pe +\h_\pa\u  +\tfrac{1}{2}\im(\uubar\cdot\hh^\pe),
\label{su.w.eq}
\\
\efac\lambda^{-1} \ww=D_x\hh^\pe +\h_\pa\uu - \i(\u\hh^\pe +\tfrac{1}{2}\h^\pe\uu),
\label{su.ww.eq}
\\
\u_t=D_x\w +\im(\uubar\cdot\ww)+\lambda\efac\h^\pe,
\label{su.u.eq}
\\
\uu_t=D_x\ww+\tfrac{1}{2} \tr(\TTh)\uu+\uu\hook\TTh+\i\u\ww-\i\w\uu +\lambda\efac\hh^\pe
\label{su.uu.eq}
\end{gather}
with the outer product notation \eqref{wedgprod}--\eqref{symmbarprod}
shown in Appendix~\ref{app:notation}.

Applying Theorem~\ref{thm:HJops} to this system \eqref{su.hpar.eq}--\eqref{su.uu.eq},
we obtain a Hamiltonian operator 
\begin{align}\label{su.Hop}
\Hop
=\bpm D_x &\im\uubar\cdot\,\\
&\\
-\i\uu&D_x+\i\u+\i\uu D_x^{-1}\im\uubar\cdot\,+\uu \rfloor D_x^{-1}\wedgbar{\uu}{}
 \epm
\end{align}
and a compatible symplectic operator 
\begin{align}\label{su.Jop}
\Jop
=\bpm \tfrac{1}{4}D_x+\u D_x^{-1}\u&\tfrac{1}{2}\im\uubar\cdot\, +\u D_x^{-1} \re\uubar\cdot\,\\
&\\
-\i\tfrac{1}{2}\uu+\uu D_x^{-1}\u & D_x-\i\u+\uu D_x^{-1}\re\uubar\cdot\,
 \epm
\end{align}
with the system being given by 
\beq\label{su.flow.system.HJ}
\bpm\u_t\\ \uu_t\epm
=\Hop\bpm\w\\\ww\epm 
+\lambda\efac \bpm\h^\pe\\\hh^\pe\epm,
\quad
\efac\lambda^{-1}\bpm\w\\\ww\epm =\Jop\bpm\h^\pe\\\hh^\pe\epm.
\eeq
Composition of these two operators yields 
\beq\label{su.flow.system.R}
\lambda\efac \bpm \u_t\\ \uu_t \epm
=\Rop\bpm \h^\pe\\\hh^\pe \epm  +\efacsq \bpm \h^\pe\\\hh^\pe \epm,
\eeq
where 
\beq\label{su.Rop}
\Rop = \Hop\Jop = \bpm \Rop_{11} & \Rop_{12} \\ \Ropbar_{12}^* & \Rop_{22} \epm
\eeq
is a hereditary recursion operator given by 
\begin{subequations}\label{su.Rop.comps}
\begin{align}
& \begin{aligned}
\Rop_{11} =&
\tfrac{1}{4} D_x^2
-\tfrac{1}{2}|\uu|^2
+D_x \u D_x^{-1}\u , 
\end{aligned}
\\
& \begin{aligned}
\Rop_{12} =&
\tfrac{1}{2}D_x\im\uubar\cdot\, 
+\im\uubar\cdot D_x 
-\u\re\uubar\cdot
+D_x\u D_x^{-1} \re\uubar\cdot\, , 
\end{aligned}
\\
%& \begin{aligned}
%\Rop_{21} =& 
%-\tfrac{1}{4}\i\uu D_x 
%-\i\tfrac{1}{2}D_x\uu
%+\tfrac{1}{2}\u\uu
%+D_x\uu D_x^{-1}\u 
%-\tfrac{1}{2}\i\uu D_x^{-1}|\uu|^2
%-\tfrac{1}{2}\i\uu\hook D_x^{-1}\symmbar{\uu}{\uu} , 
%\end{aligned}
& \begin{aligned}
\Rop_{22} =&
D_x^2 
+\u^2
-\tfrac{1}{2}\i\uu\im\uubar\cdot\,-D_x\i\u+\i \u D_x
+D_x\uu D_x^{-1}\re\uubar\cdot\,
+\i\uu D_x^{-1}\im\uubar\cdot D_x
\\
&
-\i\uu D_x^{-1}\u\re\uubar\cdot\,
+\uu\hook D_x^{-1}\wedgbar{\uu}{}D_x
-\i\uu\hook D_x^{-1}\u\symmbar{\uu}{} . 
\end{aligned}
\end{align}
\end{subequations}
Note that all operations $\re,\im,\cdot\,,\wedg{}{},\wedgbar{}{},\symm{}{},\symmbar{}{}$ are meant to act in rightmost to leftmost order. 

\begin{prop}\label{prop:SU.results}
All non-stretching curve flows $\map(x,t)$ in $SU(n+1)/U(n)$ 
are described by the system \eqref{su.flow.system.HJ} 
which is formulated by using a $U(n)$-parallel frame,
with $(\u,\uu)$ being the components of the Cartan matrix $u=\lrep(\u,\uu)\rrep$. 
This system encodes a Hamiltonian operator \eqref{su.Hop} and a compatible symplectic operator \eqref{su.Jop},
which yields a hereditary recursion operator \eqref{su.Rop}--\eqref{su.Rop.comps}. 
The action of the equivalence group $U(n-1)\subset U(n)$ of the frame 
is given by $(\u,\uu)\to (\u,\det(X_{n-1})^{-1/2}\uu X_{n-1}^{-1})$,
where $X_{n-1}$ is an arbitrary $x$-independent $(n-1)\times(n-1)$ unitary matrix. 
\end{prop}

Both operators \eqref{su.Hop} and \eqref{su.Jop} are invariant 
under $x$-translation symmetries and $U(1)$ symmetries,
which are respectively generated by 
\beq\label{su.x.symm}
\X = \u_x\hook\partial_{\u} + \uu_x\hook\partial_{\uu}
\eeq
and 
\beq\label{su.ad_j.symm}
\X = \i\uu\hook\partial_{\uu} .
\eeq

\subsection{mKdV flow}

Using the $x$-translation symmetry generator \eqref{su.x.symm},
we define a flow on $(\u,\uu)$ by taking 
\beq
(\h^\pe,\hh^\pe) =(\u_x,\uu_x) .
\eeq

This yields, from the frame structure equations \eqref{su.hpar.eq}--\eqref{su.ww.eq}, 
\begin{align}
&\h_\pa=\tfrac{1}{2}\u^2+\tfrac{1}{2}|\uu|^2,
\\
&  \lambda^{-1}\efac \TTh=\wedgbar{\uu}{\uu_x} -\tfrac{1}{2}\i\u\symmbar{\uu}{\uu},
\\
&  \lambda^{-1}\efac \w=\tfrac{1}{4}\u_{xx}+\tfrac{1}{2}(\u^2+|\uu|^2)\u +\tfrac{1}{2}\im(\uubar\cdot\uu_x),
\\
&  \lambda^{-1}\efac \ww=\uu_{xx}+\tfrac{1}{2}(\u^2+|\uu|^2)\uu -\i(\tfrac{1}{2}\u_x\uu +\u\uu_x).
\end{align}
(See Appendix~\ref{app:notation} for notation.)
Hence, the resulting flow equations \eqref{su.u.eq}--\eqref{su.uu.eq} are given by 
\begin{align}
&  \lambda^{-1}\efac \u_t -\efacsq\u_x
= \tfrac{1}{4}\u_{xxx} 
+\tfrac{3}{2}\u^2\u_x 
+\tfrac{3}{2}\im(\uubar\cdot\uu_{xx}),
\label{su.mkdv.ueq}
\\
&\begin{aligned}
 \lambda^{-1}\efac \uu_t -\efacsq \uu_x
&= \uu_{xxx}
-\tfrac{3}{4}\big( 2\i\u|\uu|^2 +6\i\im(\uubar_x\cdot\uu) +\i\u_{xx} -(\u^2)_x \big) \uu 
\\&\qquad
+\tfrac{3}{2}\big( \u^2+|\uu|^2 -\i\u_x \big) \uu_x.
\label{su.mkdv.uueq}
\end{aligned}
\end{align}
This is the integrable mKdV-type system \eqref{x.flow}. 
It has the bi-Hamiltonian structure
\beq
\bpm \u \\ \uu \epm_t 
= \Hop\bpm \delta\mk{H}/\delta\u \\ \delta\mk{H}/\delta\uu \epm 
= \Eop\bpm \delta\mk{E}/\delta\u \\ \delta\mk{E}/\delta\uu \epm,
\eeq
where the Hamiltonians are given by 
\begin{align}
&\mk{E}=\int \tfrac{1}{2}(\u^2+|\uu|^2)\, dx,
\\
&\mk{H}=\int \big( -\tfrac{1}{8}(\u_x)^2 -\tfrac{1}{2}|\uu_x|^2 +\tfrac{1}{8}(\u^2+|\uu|^2)^2 +\tfrac{1}{2}\u\im(\uubar\cdot\uu_x) \big)\, dx , 
\end{align}
and where the second Hamiltonian operator is given by 
\beq
\Eop = \Rop\Hop= \Hop\Jop\Hop 
\eeq
in terms of the recursion operator \eqref{su.Rop}.

\subsection{NLS flow}\label{sec:su.nls}

Using the $U(1)$ symmetry generator \eqref{su.ad_j.symm},
we define another flow on $(\u,\uu)$ by taking 
\beq
(\h^\pe,\hh^\pe) =(0,\i\uu) .
\eeq

This yields, from the frame structure equations \eqref{su.hpar.eq}--\eqref{su.ww.eq}, 
\begin{align}
&  \h_\pa=0,
\\
&  \lambda^{-1}\efac \w= \tfrac{1}{2} |\uu|^2,
\\
&  \lambda^{-1}\efac \ww= \i\uu_x+\u\uu,
\\
&  \lambda^{-1}\efac \TTh= \i\uubar^\t\uu=\tfrac{1}{2}\i\symmbar{\uu}{\uu},
\quad 
 \lambda^{-1}\efac \tr(\TTh)=\i|\uu|^2.
\end{align}
(See Appendix~\ref{app:notation} for notation.)
Then the resulting flow equations \eqref{su.u.eq}--\eqref{su.uu.eq} are given by 
\begin{align}
& \lambda^{-1}\efac\u_t=(|\uu|^2)_x,
\label{su.nls.ueq}
\\
& \lambda^{-1}\efac\uu_t -\efacsq\uu_x=\i\uu_{xx}+\i(\u^2+|\uu|^2)\uu + \u_x \uu,
\label{su.nls.uueq}
\end{align}
which is the integrable NLS-type system \eqref{ad_j.flow}. 
It has the bi-Hamiltonian structure
\beq
\bpm \u \\ \uu \epm_t 
= \Hop\bpm \delta\mk{H}/\delta\u \\ \delta\mk{H}/\delta\uu \epm 
= \Eop\bpm \delta\mk{E}/\delta\u \\ \delta\mk{E}/\delta\uu \epm,
\eeq
where the first Hamiltonian is given by 
\beq
\mk{H} =\int \big( \tfrac{1}{2}\u|\uu|^2 -\im(\uubar\cdot\uu_x) \big)\, dx . 
\eeq
To explain the second Hamiltonian,
we observe that the Hamiltonian operator $\Eop=\Rop\Hop$ 
will give the NLS system \eqref{su.nls.ueq}--\eqref{su.nls.uueq} if 
\beq\label{su.triv.E}
\Hop\bpm \delta\mk{E}/\delta\u \\ \delta\mk{E}/\delta\uu \epm = \bpm 0 \\ \i\uu \epm . 
\eeq
This relation can be made to hold if we take  $\mk{E}=0$ 
and allow $D_x^{-1}(0)=c=\const$ and $D_x^{-1}(\mb{0})=\CC=\const$, 
since then we have 
\beq
\Hop\bpm 0 \\ \mb{0} \epm = \bpm 0 \\ c\i\uu +\uu\hook\CC \epm 
\eeq
which reduces to the desired flow \eqref{su.triv.E} when $c=1$ and $\CC=0$. 
This type of Hamiltonian structure has been considered previously in other contexts
\cite{OlvRos,AncMob}. 

We remark that this NLS system \eqref{su.nls.ueq}--\eqref{su.nls.uueq} and its Lax pair 
has been derived previously \cite{Tsu} by applying a Miura transformation to 
a multi-component version of the Yajima-Oikawa system \cite{YajOik}.

\subsection{SG flow}

We obtain a SG-type system \eqref{kerJ.flow} by putting $\w=0$ and $\ww=0$
in the frame structure equations \eqref{su.hpar.eq}--\eqref{su.uu.eq},
which yields 
\beq\label{su.SG.flow}
\u_t=\lambda\efac\h^\pe,
\quad
\uu_t=\lambda\efac\hh^\pe
\eeq
with 
\begin{align}
& D_x\h_\pa -\h^\pe\u -\brack{\hh^\pe,\uu} =0,
\label{su.hpar.sg.eq}
\\
& D_x\h^\pe +4\h_\pa\u +2 \im(\uubar\cdot\hh^\pe)=0,
\label{su.hperp.sg.eq1}
\\
& D_x\hh^\pe +\h_\pa\uu -\i(\u\hh^\pe +\tfrac{1}{2}\h^\pe\uu)=0.
\label{su.hperp.sg.eq2}
\end{align}
These equations possess a conservation law
\beq\label{su.sg.conslaw}
D_x\big( \h_\pa^2+\tfrac{1}{4}(\h^\pe)^2+|\hh^\pe|^2 \big) =0
\eeq
from which we obtain 
\beq
\h_\pa^2+\tfrac{1}{4}(\h^\pe)^2+|\hh^\pe|^2=1
\eeq
after $t$ is conformally rescaled,
Hence we have 
\begin{align}
\h_\pa^2=\pm\sqrt{1-\tfrac{1}{4}(\h^\pe)^2+|\hh^\pe|^2} .
\end{align}
Substituting this conservation law along with the flow equations \eqref{su.SG.flow} 
into equations \eqref{su.hperp.sg.eq1}--\eqref{su.hperp.sg.eq2}, 
we get 
\begin{align}
& \u_{tx}
=\mp  4\sqrt{\efacsq\lambda^2- (\tfrac{1}{4}\u_t^2+|\uu_t|^2)}\,\u
+\im(\uubar_t\cdot\uu),
\label{su.sg.ueq}
\\
&\uu_{tx}
=\mp \sqrt{\efacsq\lambda^2-(\tfrac{1}{4}\u_t^2+|\uu_t|^2)}\,\uu
+\i(\u\uu_t+\tfrac{1}{2}\u_t\uu).
\label{su.sg.uueq}
\end{align}

\subsection{Hierarchies of integrable systems}

The mKdV system \eqref{su.mkdv.ueq}--\eqref{su.mkdv.uueq}
and the NLS system \eqref{su.nls.ueq}--\eqref{su.nls.uueq}
are each a root system in a hierarchy of integrable systems
generated by the recursion operator \eqref{su.Rop}--\eqref{su.Rop.comps}. 

\begin{thm}\label{thm:SU.hierarchy}
There is a mKdV hierarchy of integrable systems 
\beq
\bpm\u_t\\ \uu_t\epm -\lambda\efac \bpm\u_x\\\uu_x\epm 
=\Rop^k \bpm\u_x\\\uu_x\epm ,
\quad
k=0,1,2,\ldots
\eeq
as well as a NLS hierarchy of integrable systems 
\beq
\bpm\u_t\\ \uu_t\epm -\lambda\efac \bpm 0\\\i\uu\epm 
=\Rop^k \bpm 0\\\i\uu\epm ,
\quad
k=0,1,2,\ldots
\eeq
arising from the structure equations \eqref{su.hpar.eq}--\eqref{su.uu.eq} 
of a $U(n)$-parallel frame for non-stretching curve flows in $SU(n+1)/U(n)$. 
Associated to the mKdV hierarchy is an integrable SG system \eqref{su.sg.ueq}--\eqref{su.sg.uueq}. 
All of these integrable systems are invariant under the unitary symmetry group $U(n-1)$ 
which acts as $(\u,\uu_1,\uu_2)\to (\u,\uu_1X_{n-2}^{-1},\uu_2X_{n-2}^{-1})$
where $X_{n-2}$ is an arbitrary $x$-independent $(n-2)\times(n-2)$ unitary matrix. 
\end{thm}

\section{Integrable systems in $SO(2n)/U(n)$}
\label{sec:SO}

Since the symmetric space $SO(n+1)/U(n)$ has rank $[n/2] \geq 1$, 
there is a unique choice of a $U(n)$-parallel frame, up to equivalence, 
for non-stretching curve flows $\map(x,t)$ 
only in the cases $n=2,3$ when the rank is $1$ 
(see Remark~\ref{rem:equiv.frames}). 
We will choose a $U(n)$-parallel frame that does not depend on $n$. 
It is distinguished among all possible choices (when $n\geq 4$) 
by having the maximal dimension for its equivalence group. 

To construct this frame, 
we will need the following matrix decomposition of the symmetric Lie algebra $\mk{so}(2n)= \hsp\oplus\msp$:
\begin{equation}\label{so.matr.rep}
\bpm \re(\AA+\BB) &\im(\AA+\BB) \\ \im(\AA-\BB) & -\re(\AA-\BB) \epm 
\in\mk{so}(2n),
\quad 
\re\BB,\re\AA,\im\AA\in\mk{so}(n), \im\BB\in\mk{s}(n)
\end{equation}
where we will write 
\begin{align}
\lrep\BB\rrep :=& \bpm \re\BB &\im\BB \\ -\im\BB& \re\BB\epm 
\in\hsp,
\quad 
\BB\in  \mk{u}(n),
\label{so.hsp.matr.rep}
\\
\lrep\AA\rrep :=& \bpm \re\AA& \im\AA\\ \im\AA& -\re\AA\epm 
\in\msp=\mk{so}(2n)/\mk{u}(n)\simeq \Cnum^{\frac{1}{2}n(n-1)},
\quad 
\AA\in\mk{so}(n,\Cnum) . 
\label{so.msp.matr.rep}
\end{align}
In the Cartan subspace $\asp\subset\msp\simeq \Cnum^{\frac{1}{2}n(n-1)}$, 
we choose the element $\lrep\ee\wedge\ee\rrep$ given by 
$\ee\wedge\ee\in \mk{so}(n,\Cnum)$ 
with $\ee = (1,\mb{0})\in\Cnum^n = \Cnum\oplus\Cnum^{n-1}$, 
where $\mb{0}\in\Cnum^{n-1}$. 
This Cartan element $\lrep\ee\wedge\ee\rrep$ yields 
a decomposition of 
$\msp=\mpasp\oplus\mpesp$ and $\hsp=\hpasp\oplus\hpesp$ 
into centralizer subspaces and perp subspaces:
\beq
\begin{aligned}
& \AA = 
\bpm 
0&\a_\pa+\i\a_\pe &\aaone_\pe\\ 
-\a_\pa-\i\a_\pe&0&\aatwo_\pe\\
-\aaone_\pe{}^\t&-\aatwo_\pe{}^\t& \AA_\pa
\epm
\in\mk{so}(n,\Cnum),
\\& 
\AA_\pa\in\mk{so}(n-2,\Cnum),
\aaone_\pe,\aatwo_\pe\in\Cnum^{n-2},
\a_\pa,\a_\pe\in\Rnum
\end{aligned}
\label{so.msp.perp.pa.matr.rep}
\eeq
and
\beq
\begin{aligned}
& \BB = 
\bpm 
\i\bone_\pa +\i\b_\pe&\btwo_\pa& \bbone_\pe\\
-\btwobar_\pa&-\i\bone_\pa +\i\b_\pe&\bbtwo_\pe\\
-\bbonebar_\pe{}^\t&-\bbtwobar_\pe{}^\t&\BB_\pa
\epm
\in\mk{u}(n),
\\& 
\BB_\pa\in\mk{u}(n-2), 
\bbone_\pe,\bbtwo_\pe\in \Cnum^{n-2}, 
\bone_\pa,\b_\pe\in\Rnum,\btwo_\pa\in\Cnum ,
\end{aligned}
\label{so.hsp.perp.pa.matr.rep} 
\eeq
with the notation 
\begin{align}
& \lrep(\a_\pa,\AA_\pa)\rrep:=
\bpm
\bpm 0&\a_\pa &\mb{0}\\ -\a_\pa&0&\mb{0}\\ \mb{0}^\t& \mb{0}^\t& \re\AA_\pa \epm
& 
\bpm 0& 0 &\mb{0}\\ 0 &0&\mb{0}\\ \mb{0}^\t& \mb{0}^\t& \im\AA_\pa \epm
\\
\bpm 0& 0 &\mb{0}\\ 0 &0&\mb{0}\\ \mb{0}^\t& \mb{0}^\t& \im\AA_\pa \epm
& 
\bpm 0&-\a_\pa &\mb{0}\\ \a_\pa&0&\mb{0}\\ \mb{0}^\t& \mb{0}^\t& -\re\AA_\pa \epm
\epm
\in\msp_\pa
\label{so.msp.pa},
\\
& \begin{aligned}
&\lrep(\i\a_\pe,\aaone_\pe,\aatwo_\pe)\rrep:=
\\&\qquad
\bpm
\bpm 0& 0&\re\aaone_\pe\\ 0&0&\re\aatwo_\pe\\ -\re\aaone_\pe^\t& -\re\aatwo_\pe^\t&0 \epm
&
\bpm 0& \a_\pe&\im\aaone_\pe\\ -\a_\pe&0&\im\aatwo_\pe\\ -\im\aaone_\pe^\t& -\im\aatwo_\pe^\t&0 \epm
\\
\bpm 0& \a_\pe&\im\aaone_\pe\\ -\a_\pe&0&\im\aatwo_\pe\\ -\im\aaone_\pe^\t& -\im\aatwo_\pe^\t&0 \epm
&
\bpm 0& 0& -\re\aaone_\pe\\ 0&0& -\re\aatwo_\pe\\ \re\aaone_\pe^\t& \re\aatwo_\pe^\t&0 \epm
\epm
\in\msp_\pe,
\end{aligned}
\label{so.msp.pe}
\end{align}
and
\begin{align}
& \begin{aligned}
&\lrep(\i\bone_\pa,\btwo_\pa,\BB_\pa)\rrep :=
\\&\qquad
\bpm
\bpm 0&\re\btwo_\pa&0\\-\re\btwobar_\pa&0&0\\0&0&\re\BB_\pa\epm
& 
\bpm \bone_\pa&\im\btwo_\pa&0\\-\im\btwobar_\pa&-\bone_\pa&0\\0&0&\im\BB_\pa\epm
\\
\bpm -\bone_\pa&-\im\btwo_\pa&0\\\im\btwobar_\pa&\bone_\pa&0\\0&0&-\im\BB_\pa\epm
&
\bpm 0&\re\btwo_\pa&0\\-\re\btwobar_\pa&0&0\\0&0&\re\BB_\pa\epm
\epm
\in\hsp_\pa ,
\end{aligned}
\label{so.hsp.pa}
\\
& \begin{aligned}
& \lrep(\i\b_\pe,\bbone_\pe,\bbtwo_\pe)\rrep :=
\\&\quad
\bpm
\bpm 0&0&\re\bbone_\pe\\ 0&0&\re\bbtwo_\pe\\ -\re\bbonebar_\pe{}^\t&-\re\bbtwobar_\pe{}^\t&0\epm
& 
\bpm \b_\pe&0&\im\bbone_\pe\\ 0&\b_\pe&\im\bbtwo_\pe\\ -\im\bbonebar_\pe{}^\t&-\im\bbtwobar_\pe{}^\t&0\epm
\\
\bpm -\b_\pe&0&-\im\bbone_\pe\\ 0&-\b_\pe&-\im\bbtwo_\pe\\ \im\bbonebar_\pe{}^\t&\im\bbtwobar_\pe{}^\t&0\epm
& 
\bpm 0&0&\re\bbone_\pe\\ 0&0&\re\bbtwo_\pe\\ -\re\bbonebar_\pe{}^\t&-\re\bbtwobar_\pe{}^\t&0\epm
\epm
\in\hsp_\pe.
\end{aligned}
\label{so.hsp.pe}
\end{align}
Details of this decomposition and its algebraic properties, 
including the Lie bracket structure, 
are summarized in Appendix~\ref{app:so.liealg}. 

We now write out the frame structure equations \eqref{cartaneq.mpar}--\eqref{cartaneq.hperp}
by using the matrix representations \eqref{so.msp.pa}--\eqref{so.hsp.pe}. 

First, the tangential components \eqref{pull.tangential} of 
the linear coframe $e$ and the linear connection $\conx$ 
along the curve are given by the variables
\begin{align}
&
\e=\efac \lrep(1,\mb{0})\rrep 
\in\msp_\pa\subset\msp,
\label{so.e}
\\
&
u=\lrep(\i\u,\uu_1,\uu_2)\rrep
\in\i\Rnum\oplus\Cnum^{n-2}\oplus\Cnum^{n-2}\simeq \hsp_\pe,
\label{so.u}
\end{align}
where $\u\in\Rnum$ is a real scalar variable, 
and $\uu_1,\uu_2\in\Cnum^{n-1}$ are complex vector variables. 
These variables geometrically represent the components of the Cartan matrix of the $U(n)$-parallel frame. 
Here $\norme=4\sqrt{n-1}$ is a normalization constant determined by the Cartan-Killing metric 
(see Proposition~\ref{prop:so.struct} in Appendix~\ref{app:so.liealg}). 

Next, the flow components \eqref{pull.flow} of 
the linear coframe and the linear connection 
are expressed in terms of the variables
\begin{align}
&
h_\pa=\lambda\lrep(\h_\pa,\HH_\pa)\rrep
\in\Rnum\oplus\mk{so}(n-2,\Cnum)\simeq\msp_\pa,
\label{so.h.par}
\\
&
h_\pe=\lrep(\i\h_\pe,\hh_{1\pe},\hh_{2\pe})\rrep
\in\i\Rnum\oplus\Cnum^{n-2}\oplus\Cnum^{n-2}\simeq \msp_\pe,
\label{so.h.perp}
\\
&
\varpi^\pa=\lrep(\i\thone,\thtwo,\TTh)\rrep
\in\i\Rnum\oplus\Cnum\oplus\mk{u}(n-2)\simeq \hsp_\pa,
\label{so.w.par}
\\
& 
\varpi^\pe=\lrep(\i\w,\ww_1,\ww_2)\rrep
\in\i\Rnum\oplus\Cnum^{n-2}\oplus\Cnum^{n-2}\simeq \hsp_\pe,
\label{so.w.perp}
\end{align}
as well as 
\beq
h^\pe = \lambda\efac\lrep(\i\h^\pe,\hh^{1\pe},\hh^{2\pe})\rrep
\in\hsp_\pe
\label{so.h.perp.up}
%\ad(\e)h_\pe =\efac \lrep(-2\i\h_\pe,\hh_{2\pe},-\hh_{1\pe})\rrep
\eeq
from relation \eqref{hperp.up}. 
Here
$\h_\pa$, $\h_\pe$, $\h^\pe$, $\thone$, $\w$ $\in\Rnum$ are real scalar variables; 
$\thtwo$ is a complex scalar variable;
$\hh_{1\pe}$, $\hh^{1\pe}$, $\hh_{2\pe}$, $\hh^{2\pe}$, $\ww_1$, $\ww_2$ $\in\Cnum^{n-2}$ are complex vector variables; 
$\HH_\pa$ $\in\mk{so}(n-2)$ is a complex antisymmetric matrix variable;
$\TTh$ $\in\mk{su}(n-2)$ is a unitary matrix variable.

Then, the frame structure equations \eqref{cartaneq.mpar}--\eqref{cartaneq.hperp}
are given by 
\begin{align}
&
D_x\h_\pa = \h^\pe\u +\brack{\uu_1,\hh^{1\pe}} +\brack{\uu_2,\hh^{2\pe}},
\label{so.hpar.eq}
\\
&
D_x\HH_\pa = \wedg{\uu_2}{\hh^{1\pe}}-\wedg{\uu_1}{\hh^{2\pe}},
\label{so.HHpar.eq}
\\
&
\lambda^{-1}\efac \w= 
\tfrac{1}{4}D_x\h^\pe  +\h_\pa\u 
+\tfrac{1}{2}\im(\uubar_1\cdot\hh^{1\pe}+\uubar_2\cdot\hh^{2\pe}), 
\label{so.w.eq}
\\
&
\lambda^{-1}\efac \ww_1
=D_x\hh^{1\pe}+\h_\pa\uu_1+\uubar_2\hook\HH_\pa -\i(\tfrac{1}{2}\h^\pe\uu_1+\u\hh^{1\pe}),
\label{so.ww1.eq}
\\
&
\lambda^{-1}\efac \ww_2
=D_x\hh^{2\pe}+\h_\pa\uu_2 -\uubar_1\hook\HH_\pa  -\i(\tfrac{1}{2}\h^\pe\uu_2 +\u\hh^{2\pe})
\label{so.ww2.eq}
\end{align}
and
\begin{align}
&
D_x\thone = \im(\wwbar_1\cdot\uu_1-\wwbar_2\cdot\uu_1),
\label{so.thone.eq}
\\
&
D_x\thtwo = \wwbar_2\cdot\uu_1 -\uubar_2\cdot\ww_1, 
\label{so.thtwo.eq}
\\
&
D_x\TTh = \wedgbar{\uu_1}{\ww_1}+\wedgbar{\uu_2}{\ww_2},
\label{so.TTh.eq}
\\
&
\u_{t}
=D_x\w +\im(\uubar_1\cdot\ww_1+\uubar_2\cdot\ww_2)
+\lambda\efac \h^\pe,
\label{so.u.eq}
\\
&
\uu_{1 t}
=D_x\ww_1 +\uu_1\hook\TTh -\i\thone\uu_{1}-\thtwo\uu_2 +\i(\u\ww_1-\w\uu_1)
+ \lambda\efac \hh^{1\pe},
\label{so.uu1.eq}
\\
&
\uu_{2 t}
=D_x\ww_2 + \uu_2\hook\TTh +\i\thone\uu_{2}+\thtwobar\uu_1  +\i(\u\ww_2-\w\uu_2) 
+\lambda\efac \hh^{2\pe},
\label{so.uu2.eq}
\end{align}
with the outer product notation \eqref{wedgprod}--\eqref{symmbarprod}
shown in Appendix~\ref{app:notation}.

We apply Theorem~\ref{thm:HJops} to this system \eqref{so.hpar.eq}--\eqref{so.uu2.eq},
yielding a Hamiltonian operator 
\beq\label{so.Hop}
\Hop
=\bpm 
D_x&\im\uubar_1\cdot\,&\im\uubar_2\cdot\,\\
&&\\
-\i\uu_1&D_x+\i\u-\uu_2D_x^{-1}\uubar_2\cdot\,&\uu_2 D_x^{-1}\uu_1\cdot\, \mathcal{C}
\\
&-\i\uu_1 D_x^{-1}\im\uubar_1\cdot\,+\uu_1\hook D_x^{-1}\wedgbar{\uu_1}{}&
+\i\uu_1 D_x^{-1}\im\uubar_2\cdot\,+\uu_1\hook D_x^{-1}\wedgbar{\uu_2}{}
\\
&&
\\
-\i\uu_2&-\uu_1D_x^{-1}\uu_2\cdot\, \mathcal{C}&D_x+\i\u+\uu_1 D_x^{-1}\uubar_1\cdot\, 
\\
&-\i\uu_2 D_x^{-1}\im(\uubar_1\cdot\,)+\uu_2\hook D_x^{-1}\wedgbar{\uu_1}{}&
+\i\uu_2 D_x^{-1}\im(\uubar_2\cdot\,)+\uu_2\hook D_x^{-1}\wedgbar{\uu_2}{}
 \epm
\eeq
and a compatible symplectic operator 
\beq\label{so.Jop}
\Jop=\bpm
\tfrac{1}{4}D_x+\u D_x^{-1}\u
&
\tfrac{1}{2}\im\uubar_1\cdot\, +\u D_x^{-1} \re\uubar_1\cdot\,
&
\tfrac{1}{2}\im\uubar_2\cdot\,+\u D_x^{-1} \re\uubar_2\cdot\,
\\
&&\\
-\i\tfrac{1}{2}\uu_1+\uu_1 D_x^{-1}\u 
& D_x-\i\u
&\uu_1D_x^{-1}\re\uubar_2\cdot\, -\uubar_2\hook D_x^{-1}\uu_1\wedge
\\
&+\uu_1 D_x^{-1}\re\uubar_1\cdot\,+\uubar_2\hook D_x^{-1}\uu_2\wedge&
\\
&&\\
-\i\tfrac{1}{2}\uu_2+\uu_2 D_x^{-1}\u 
& \uu_2 D_x^{-1}\re\uubar_1\cdot\, -\uubar_1\hook D_x^{-1}\uu_2\wedge
& D_x-\i\u
\\
&&+\uu_2D_x^{-1}\re\uubar_2\cdot\, +\uubar_1\hook D_x^{-1}\uu_1\wedge
 \epm,
\eeq
where $\mathcal{C}$ denotes the complex conjugation operator, 
with the system being given by 
\beq\label{so.flow.system.HJ}
\bpm\u_t\\ \uu_{1t}\\\uu_{1t}\epm
=\Hop\bpm\w\\\ww_1\\\ww_2\epm
+\lambda\efac\bpm\h^\pe\\\hh^{1\pe}\\\hh^{2\pe}\epm,
\quad
\efac\lambda^{-1}\bpm\w\\\ww_1\\\ww_2\epm
=\Jop\bpm\h^\pe\\\hh^{1\pe}\\\hh^{2\pe}\epm.
\eeq
Composition of these two operators yields 
\beq\label{so.flow.system.R}
\bpm \u_t\\ \uu_{1t}\\\uu_{1t} \epm
=\Rop\bpm \w\\\ww_1\\\ww_2 \epm
+\lambda\efac\bpm \h^\pe\\\hh^{1\pe}\\\hh^{2\pe} \epm,
\eeq
where 
\beq\label{so.Rop}
\Rop = \Hop\Jop =\bpm \Rop_{11}& \Rop_{12}&\Rop_{13}\\ \Ropbar_{12}^*&\Rop_{22}&\Rop_{23} \\\Ropbar_{13}^* &\Ropbar_{23}^* &\Rop_{33}
 \epm
\eeq
is a hereditary recursion operator given by 
\begin{align}
& \begin{aligned}
\Rop_{11}&=
\tfrac{1}{4}D_x^2
+\u^2
-\tfrac{1}{2}(|\uu_1|^2+|\uu_2|^2)
+D_x\u D_x^{-1}\u,
\end{aligned}
\\
& \begin{aligned}
R_{12}&=
\tfrac{1}{2}D_x\im\uubar_1\cdot
+D_x\u  D_x^{-1}\re\uubar_1\cdot
+\im\uubar_1\cdot D_x
-\u\re\uubar_1\cdot
-2\im \uubar_2\cdot\uubar_1\rfloor D_x^{-1}\uu_2\wedge,
\end{aligned}
\\
& \begin{aligned}
R_{13}&=
\tfrac{1}{2}D_x\im\uubar_2\cdot
+D_x\u D_x^{-1}\re\uubar_2\cdot
+\im\uubar_2\cdot D_x
-\u\re\uubar_2\cdot
-2\im\uubar_1\cdot\uubar_2\hook D_x^{-1}\uu_1\wedge,
\end{aligned}
\\
& \begin{aligned}
R_{22}&=
D_x^2 +\i(\u D_x-D_x\u)
+\u^2
-\tfrac{1}{2}\i\uu_1\im\uubar_1\cdot
\\&\qquad
+D_x\uu_{1}D_x^{-1}\re\uubar_1\cdot
+D_x\uubar_2\hook D_x^{-1}\uu_2\wedge
+\i\u\uubar_2 D_x^{-1}\uu_2\wedge
\\&\qquad
-\uu_2 D_x^{-1}\uubar_2\cdot D_x
%-\uu_2 D_x^{-1} \uubar_2\cdot\uubar_2\hook D_x^{-1}\uu_2\wedge
-\i\uu_1 D_x^{-1}\im\uubar_1\cdot D_x
+\uu_1\hook D_x^{-1}\wedgbar{\uu_1}{}D_x
%-\i\uu_1 D_x^{-1}\im \uubar_1\cdot\uubar_2\hook D_x^{-1}\uu_2\wedge
\\&\qquad
+\i\uu_2 D_x^{-1}\u\uubar_2\cdot
+\i\uu_1 D_x^{-1}\u\re\uubar_1\cdot
-\i\uu_1 \hook D_x^{-1}\u\symmbar{\uu_1}{}
\\&\qquad
+\uu_1 D_x^{-1}\wedgbar{\uu_1}{\uubar_2\hook D_x^{-1}\uu_2\wedge}
%-\uu_2 D_x^{-1}\uu_1\cdot\mathcal{C}\uubar_1 \hook D_x^{-1}\uu_2\wedge
%-\i\uu_1 D_x^{-1}\im\uubar_2\cdot\uubar_1\hook D_x^{-1}\uu_2\wedge 
-\uu_1\hook D_x^{-1}\wedgbar{\uu_2}{\uubar_1\hook D_x^{-1}\uu_2\wedge},
\end{aligned}
\\
& \begin{aligned}
R_{23}&=
-\tfrac{1}{2}\i\uu_1\im\uubar_2\cdot
+D_x\uu_1 D_x^{-1}\re\uubar_2\cdot-D_x\uubar_2\hook D_x^{-1}\uu_1\wedge
-\i\u\uubar_2\hook D_x^{-1}\uu_1\wedge
\\&\qquad
%-\uu_2 D_x^{-1}\uubar_2\cdot\uu_1 D_x^{-1}\re\uubar_2\cdot
%+\uu_2 D_x^{-1}\uubar_2\cdot\uubar_2\hook D_x^{-1}\uu_1\wedge
%+\i\uu_1 D_x^{-1}\im\uubar_1\cdot\uubar_2\hook D_x^{-1}\uu_1\wedge
+\uu_2 D_x^{-1}\uu_1\cdot\mathcal{C} D_x
+\uu_1 D_x^{-1}\wedgbar{\uu_2}{D_x}
%+\uu_2 D_x^{-1}\uu_1\cdot\uubar_2 D_x^{-1}\re\uubar_2\cdot
%+\uu_2D_x^{-1}\uu_1\cdot \mathcal{C}\uubar_1\hook D_x^{-1}\uu_1\wedge
+\i\uu_1 D_x^{-1}\im\uubar_2\cdot D_x
\\&\qquad
-\i\uu_1 D_x^{-1}\u \re\uubar_2\cdot
+\i\uu_2D_x^{-1}\u\uu_1\cdot\mathcal{C}
%+\i\uu_1D_x^{-1}\im\uubar_2\cdot\uubar_1 \hook D_x^{-1}\uu_1\wedge
-\i\uu_1\rfloor D_x^{-1}\u\symmbar{\uu_2}{}
\\&\qquad
-\uu_1 D_x^{-1}\wedgbar{\uu_1}{\uubar_2\hook D_x^{-1}\uu_1\wedge}
+\uu_1 D_x^{-1}\wedgbar{\uu_2}{\uubar_1 \hook D_x^{-1}\uu_1\wedge},
\end{aligned}
\\
& \begin{aligned}
R_{33}&=
D_x^2+\i(\u D_x -D_x\u)
+\u^2
-\tfrac{1}{2}\i\uu_2\im\uubar_2\cdot
%+\uu_1D_x^{-1}\uu_2\cdot\mathcal{C}\uubar_2\hook D_x^{-1}\uu_1\wedge
%+\i\uu_2 D_x^{-1}\im\uubar_1\cdot\uubar_2 \hook D_x^{-1}\uu_1\wedge
\\&\qquad
+D_x\uu_2D_x^{-1}\re\uubar_2\cdot
+D_x\uubar_1\rfloor D_x^{-1}\uu_1\wedge
+\i\u\uubar_1\rfloor D_x^{-1}\uu_1\wedge
\\&\qquad
+\uu_1 D_x^{-1}\uubar_1\cdot D_x
%+\uu_1 D_x^{-1}\uubar_1\cdot\uubar_1 \hook D_x^{-1}\uu_1\wedge
+\i\uu_2 D_x^{-1}\im\uubar_2\cdot D_x
+\uu_2 D_x^{-1}\wedgbar{\uu_2}{D_x}
\\&\qquad
-\i\uu_1 D_x^{-1}\u\uubar_1\cdot
-\i\uu_2 D_x^{-1}\u\re\uubar_2\cdot 
%+\i\uu_2D_x^{-1}\im\uubar_2\cdot\uubar_1\rfloor D_x^{-1}\uu_1\wedge 
-\i\uu_2\hook D_x^{-1}\u\symmbar{\uu_2}{}
\\&\qquad
-\uu_2 D_x^{-1}\wedgbar{\uu_1}{\uubar_2\hook D_x^{-1}\uu_1\wedge}
+\uu_2 D_x^{-1}\wedgbar{\uu_2}{\uubar_1 \rfloor D_x^{-1}\uu_1\wedge }.
\end{aligned}
\end{align}
Note that all operations $\re,\im,\cdot\,,\wedg{}{},\wedgbar{}{},\symm{}{},\symmbar{}{}$ are meant to act in rightmost to leftmost order. 

\begin{prop}\label{prop:SO.results}
All non-stretching curve flows $\map(x,t)$ in $SO(2n)/U(n)$ 
are described by the system \eqref{so.flow.system.HJ} 
which is formulated by using a $U(n)$-parallel frame 
whose equivalence group has maximal size,
with $(\u,\uu_1,\uu_2)$ being the components of the Cartan matrix $u=\lrep(\u,\uu_1,\uu_2)\rrep$. 
This system encodes a Hamiltonian operator \eqref{so.Hop} and a compatible symplectic operator \eqref{so.Jop},
which yields a hereditary recursion operator \eqref{so.Rop}. 
The action of the unitary subgroup $U(n-2)$ in the equivalence group $U(n-2)\times SU(2)\subset U(n)$ of the frame 
is given by $(\u,\uu_1,\uu_2)\to (\u,\uu_1X_{n-2}^{-1},\uu_2X_{n-2}^{-1})$
where $X_{n-2}$ is an arbitrary $x$-independent $(n-2)\times(n-2)$unitary matrix. 
\end{prop}

Both operators \eqref{su.Hop} and \eqref{su.Jop} are invariant 
under $x$-translation symmetries and $U(1)$ symmetries,
which are respectively generated by 
\beq\label{so.x.symm}
\X = \u_x\hook\partial_{\u} + \uu_{1x}\hook\partial_{\uu_1}+ \uu_{2x}\hook\partial_{\uu_2}
\eeq
and 
\beq\label{so.ad_j.symm}
\X = \i\uu_1\hook\partial_{\uu_1} + \i\uu_2\hook\partial_{\uu_2} . 
\eeq

\subsection{mKdV flow}

Using the $x$-translation symmetry generator \eqref{so.x.symm},
we define a flow on $(\u,\uu_1,\uu_2)$ by taking 
\beq
(\h^\pe,\hh^{1\pe},\hh^{2\pe}) =(\u_x,\uu_{1x},\uu_{2x}) .
\eeq

The frame structure equations \eqref{so.hpar.eq}--\eqref{so.TTh.eq}
then yield 
\begin{align}
&\h_\pa=\tfrac{1}{2} (\u^2+|\uu_1|^2+|\uu_2|^2), 
\\
&\HH_\pa=\wedg{\uu_2}{\uu_1},
\\
& \lambda^{-1}\efac\w=
\tfrac{1}{4} \u_{xx}
+\tfrac{1}{2}(\u^2+|\uu_1|^2+|\uu_2|^2)\u 
+\tfrac{1}{2}\im(\uubar_1\cdot \uu_{1x} +\uubar_2\cdot \uu_{2x}),
\\
& \lambda^{-1}\efac\ww_1 =
\uu_{1xx}+\tfrac{1}{2}(\u^2+|\uu_1|^2+|\uu_2|^2)\uu_1
+|\uu_2|^2\uu_1 - (\uubar_2\cdot \uu_1)\uu_2
-\i(\tfrac{1}{2}\u_x\uu_1+\u\uu_{1x}),
\\
& \lambda^{-1}\efac\ww_2
=\uu_{2xx}+\tfrac{1}{2}(\u^2+|\uu_1|^2+|\uu_2|^2)\uu_2
+|\uu_1|^2\uu_2 - (\uubar_1\cdot \uu_2)\uu_1
-\i(\tfrac{1}{2}\u_x\uu_{2} +\u\uu_{2x})
\end{align}
and
\begin{align}
& \lambda^{-1}\efac\i\thone=
-\tfrac{1}{2}\i\u(|\uu_2|^2-|\uu_1|^2)
+\tfrac{1}{2}\tr(\wedgbar{\uu_2}{\uu_{2x}}-\wedgbar{\uu_1}{\uu_{1x}}),
\\
& \lambda^{-1}\efac\thtwo=
\uubar_{2x}\cdot\uu_1-\uubar_2\cdot\uu_{1x}
+\i\u\uubar_2\cdot\uu_1,
\\
& \lambda^{-1}\efac\TTh=
\wedgbar{\uu_1}{\uu_{1x}}+\wedgbar{\uu_2}{\uu_{2x}}
-\i\u(\uubar_1^\t\uu_1+\uubar_2^\t\uu_2).
\end{align}
(See Appendix~\ref{app:notation} for notation.)
Hence, the resulting flow equations \eqref{so.u.eq}--\eqref{so.uu2.eq} are given by 
\begin{align}
& \efac \lambda^{-1}\u_t -\efacsq\u_x
= \tfrac{1}{4}\u_{xxx}+\tfrac{3}{2}\u^2\u_x
+\tfrac{3}{2}\im(\uubar_1\cdot\uu_{1xx}+\uubar_2\cdot\uu_{2x}),
\label{so.mkdv.ueq}
\\
&\begin{aligned}
\efac\lambda^{-1}\uu_{1\,t}-\efacsq\uu_{1x}
& = \uu_{1xxx}+\big(\tfrac{3}{2}(\u^2+|\uu_1|^2+|\uu_2|^2)-\tfrac{3}{2}\i\u_x\big)\uu_{1x}
-3(\i\u\uubar_2\cdot\uu_1 +\uubar_{2x}\cdot\uu_1\big)\uu_2
\\&\qquad
+\tfrac{3}{4}\big( 2\i\u(|\uu_2|^2-|\uu_1|^2)+2\u\u_x -\i\u_{xx}
-2\i\im(\uubar_{1x}\cdot\uu_1)
\\&\qquad
+3\uubar_{2x}\cdot\uu_2+\uubar_2\cdot\uu_{2x}\big)\uu_1,
\end{aligned}
\label{so.mkdv.u1eq}
\\
&\begin{aligned}
\efac\lambda^{-1}\uu_{2\,t}-\efacsq\uu_{2x}
& = \uu_{2xxx}+\big(\tfrac{3}{2}(\u^2+|\uu_1|^2+|\uu_2|^2)-\tfrac{3}{2}\i\u_x\big)\uu_{2x}
-3(\i\u\uubar_1\cdot\uu_2 +\uubar_{1x}\cdot\uu_2)\uu_1
\\&\qquad
-\tfrac{3}{4}\big( 2\i\u(|\uu_2|^2-|\uu_1|^2) -2\u\u_x +\i\u_{xx}
+2\i\im(\uubar_{2x}\cdot\uu_2) 
\\&\qquad
-3\uubar_{1x}\cdot\uu_1 -\uubar_1\cdot\uu_{1x}\big)\uu_2.
\end{aligned}
\label{so.mkdv.u2eq}
\end{align}
This is the integrable mKdV-type system \eqref{x.flow}. 
It has the bi-Hamiltonian structure
\beq
\bpm \u \\ \uu_1 \\ \uu_2 \epm_t 
= \Hop\bpm \delta\mk{H}/\delta\u \\ \delta\mk{H}/\delta\uu_1 \\ \delta\mk{H}/\delta\uu_2 \epm 
= \Eop\bpm \delta\mk{E}/\delta\u \\ \delta\mk{E}/\delta\uu_1\\ \delta\mk{E}/\delta\uu_2
 \epm,
\eeq
where the first Hamiltonian is given by 
\beq
\mk{H} = \int \tfrac{1}{2}(\u^2+|\uu_1|^2+|\uu_2|^2)\, dx 
\eeq
and where the second Hamiltonian operator is given by 
\beq
\Eop = \Rop\Hop= \Hop\Jop\Hop 
\eeq
in terms of the recursion operator \eqref{so.Rop}.

\subsection{NLS flow}

Using the $U(1)$ symmetry generator \eqref{so.ad_j.symm},
we define another flow on $(\u,\uu_1,\uu_2)$ by taking 
\beq
(\h^\pe,\hh^{1\pe},\hh^{2\pe}) =(0,\i\uu_1,\i\uu_2) .
\eeq

This yields, from the frame structure equations \eqref{so.hpar.eq}--\eqref{so.TTh.eq},
\begin{align}
&\h_\pa=0,
\\
&\HH_\pa=2\i D_x^{-1}(\uu_2\wedge\uu_1),
\\
&\lambda^{-1}\efac\w=\tfrac{1}{2}(|\uu_1|^2+|\uu_2|^2),
\\
&\lambda^{-1}\efac\ww_1=\i\uu_{1x}+2\i\uubar_2 \hook D_x^{-1}(\uu_2\wedge\uu_1)+\u\uu_1,
\\
&\lambda^{-1}\efac\ww_2=\i\uu_{2x}-2\i\uubar_1\hook D_x^{-1}(\uu_2\wedge\uu_1)+\u\uu_2
\end{align}
and hence 
\begin{align}
\\
&\lambda^{-1}\efac\i\thone
=\tfrac{1}{2}\i(|\uu_2|^2-|\uu_1|^2),
\\
&\lambda^{-1}\efac\thtwo
=-\i\uubar_2\cdot\uu_1,
\\
&\lambda^{-1}\efac\TTh
=\i\tfrac{1}{2}(\symmbar{\uu_1}{\uu_1}+\symmbar{\uu_2}{\uu_2}).
\end{align}
(See Appendix~\ref{app:notation} for notation.)
Then the resulting flow equations \eqref{so.u.eq}--\eqref{so.uu2.eq} are given by 
\begin{align}
& \lambda^{-1}\efac\u_t
=(|\uu_1|^2+|\uu_2|^2+|D_x^{-1}(\uu_2\wedge\uu_1)|^2)_x,
\label{so.nls.ueq}
\\
& \lambda^{-1}\efac\uu_{1 t} -\efacsq\uu_{1 x}
=\i\uu_{1xx}+\uu_1\u_x+\i(\u^2+|\uu_1|^2+|\uu_2|^2)\uu_1+2(\i\uubar_{2x} -\u\uubar_2)\hook D_x^{-1}(\uu_2\wedge\uu_1),
\label{so.nls.u1eq}
\\
& \lambda^{-1}\efac\uu_{2 t} -\efacsq\uu_{2 x}
=\i\uu_{2xx}+\uu_2\u_x+\i(\u^2+|\uu_1|^2+|\uu_2|^2)\uu_2  -2(\i\uubar_{1x} -\u\uubar_1)\hook D_x^{-1}(\uu_2\wedge\uu_1),
\label{so.nls.u2eq}
\end{align}
which is the integrable NLS-type system \eqref{ad_j.flow}. 
It has the bi-Hamiltonian structure
\beq
\bpm \u \\ \uu_1 \\ \uu_2 \epm_t 
= \Hop\bpm \delta\mk{H}/\delta\u \\ \delta\mk{H}/\delta\uu_1 \\ \delta\mk{H}/\delta\uu_2 \epm 
= \Eop\bpm \delta\mk{E}/\delta\u \\ \delta\mk{E}/\delta\uu_1 \\ \delta\mk{E}/\delta\uu_2 \epm,
\eeq
where the first Hamiltonian is given by 
\begin{align}
\mk{H}=\int \tfrac{1}{2}\big( \u(|\uu_1|^2+|\uu_2|^2) -\im(\uubar_1\cdot\uu_{1x}+\uubar_2\cdot\uu_{2x}) +\im((\uubar_2\wedge\uubar_1)\cdot  D_x^{-1}(\uu_2\wedge\uu_1)) \big)\, dx .
\end{align}
The second Hamiltonian involves the same explanation that was 
given for the NLS system obtained in \secref{sec:su.nls}. 
We observe that the Hamiltonian operator $\Eop=\Rop\Hop$ 
will give the NLS system here if 
\beq\label{so.triv.E}
\Hop\bpm \delta\mk{E}/\delta\u \\ \delta\mk{E}/\delta\uu \epm = \bpm 0 \\ i\uu_1 \\ i\uu_2 \epm . 
\eeq
To make this relation hold, we take  $\mk{E}=0$ 
and allow $D_x^{-1}(0)=c=\const$ and $D_x^{-1}(\mb{0})=\CC=\const$. 
This yields 
\beq
\Hop\bpm 0 \\ \mb{0} \\ \mb{0} \epm = 
\bpm 0 \\ c_1\i\uu_1 +c_2 \uu_2 + \uu_1\hook\CC_1 \\ c_3\i\uu_2 +c_4\uu_1 + \uu_2\hook\CC_2 \epm , 
\eeq
which reduces to the desired flow \eqref{so.triv.E} when $c_1=c_3=1$, $c_2=c_4=0$, and $\CC_1=\CC_2=0$.

\subsection{SG flow}

We obtain a SG-type system \eqref{kerJ.flow} by putting $\w=0$ and $\ww_1=\ww_2=0$
in the frame structure equations \eqref{so.hpar.eq}--\eqref{so.uu2.eq}.
This yields 
\beq\label{so.sg.flow}
\u_{t}=\lambda\efac \h^\pe,
\quad
\uu_{1 t}= \lambda\efac \hh^{1\pe},
\quad
\uu_{2 t}=\lambda\efac \hh^{2\pe}
\eeq
with 
\begin{align}
&
D_x\h_\pa-\h^\pe\u -(\brack{\uu_1,\hh^{1\pe}} +\brack{\uu_2,\hh^{2\pe}})=0,
\label{so.hpar.sg.eq}
\\
&
D_x\HH_\pa +\wedg{\uu_1}{\hh^{2\pe}}-\wedg{\uu_2}{\hh^{1\pe}}=0,
\label{so.Hpar.sg.eq}
\\
&
D_x\h^\pe  +4\h_\pa\u +\im(\uubar_1\cdot\hh^{1\pe}+\uubar_2\cdot\hh^{2\pe})=0,
\label{so.hperp.sg.eq}
\\
&
D_x\hh^{1\pe}+(\h_\pa\uu_1+\uubar_2\HH_\pa)-\i(\tfrac{1}{2}\h^\pe\uu_1+\u\hh^{1\pe})=0,
\label{so.hperp.sg.eq1}
\\
&
D_x\hh^{2\pe}+(\h_\pa\uu_2 -\uubar_1\HH_\pa) -\i(\tfrac{1}{2}\h^\pe\uu_2 +\u\hh^{2\pe})=0.
\label{so.hperp.sg.eq2}
\end{align}
These equations possess a conservation law
\beq
D_x\big( \tfrac{1}{4}(\h^\pe)^2+|\hh^{1\pe}|^2+|\hh^{2\pe}|^2 +\h_\pa^2+ \tfrac{1}{2}\tr(\HHbar_\pa\HH_\pa) \big) =0
\eeq
from which we get 
\beq\label{so.sg.conslaw}
\tfrac{1}{4}(\h^\pe)^2+|\hh^{1\pe}|^2+|\hh^{2\pe}|^2 +\h_\pa^2+ \tfrac{1}{2}|\HH_\pa|^2 =1
\eeq
after $t$ is conformally rescaled. 
Hence we obtain 
\beq
\h_\pa^2 = 1- \tfrac{1}{4}(\h^\pe)^2-|\hh^{1\pe}|^2-|\hh^{2\pe}|^2 -\tfrac{1}{2}|\HH_\pa|^2 .
\eeq
By substituting this conservation law along with the flow equations \eqref{so.sg.flow} 
into equations \eqref{so.hperp.sg.eq}--\eqref{so.hperp.sg.eq2}, 
we get a nonlocal SG-type system 
\begin{align}
& \u_{tx} +4\tilde\h_\pa\u +\im(\uubar_1\cdot\uu_{1 t}+\uubar_2\cdot\uu_{2 t})=0,
\label{so.sg.ueq}
\\
& \uu_{1 tx} +(\tilde\h_\pa\uu_1+\uubar_2\hook\tilde\HH_\pa)-\i(\tfrac{1}{2}\u_t\uu_1+\u\uu_{1 t})=0,
\label{so.sg.u1eq}
\\
& \uu_{2 tx}+(\tilde\h_\pa\uu_2 -\uubar_1\hook\tilde\HH_\pa) -\i(\tfrac{1}{2}\u_t\uu_2 +\u\uu_{2 t})=0,
\label{so.sg.u2eq}
\end{align}
where
\begin{align}
& \tilde\h_\pa = \pm\sqrt{\efacsq\lambda^2- \tfrac{1}{4}\u^2-|\uu_{1 t}|^2-|\uu_{2 t}|^2 -\tfrac{1}{2}|\HH_\pa|^2},
\\
& \tilde\HH_\pa = D_x^{-1}\big( \wedg{\uu_2}{\uu_{1 t}}-\wedg{\uu_1}{\uu_{2 t}} \big).
\end{align}

We can obtain a local reduction of this SG system by the following ansatz. 
Put 
\beq\label{so.H.par}
\HH_\pa=\alpha(\h_\pa,\h^\pe)\wedg{\hh^{2\pe}}{\hh^{1\pe}}
\eeq
with $\alpha$ taken to be an unknown function of $(\h_\pa,\h^\pe)$. 
This form for $\alpha$ is motivated by the property that both $\h_\pa$ and $\h^\pe$ 
are invariant under the frame equivalence group (see Lemma~\ref{lem:su.equiv.group}). 
We can now determine $\alpha(\h_\pa,\h^\pe)$ 
by substituting this ansatz \eqref{so.H.par} 
into equations \eqref{so.hpar.sg.eq}--\eqref{so.hperp.sg.eq2} 
to get a system of linear first-order PDEs which are readily solved. 
Omitting the details, we find 
\beq 
\alpha=1/(\tfrac{1}{2}\i\h^\pe-\h_\pa)
\eeq
and hence
\beq
|\alpha|^{-2} = \tfrac{1}{4}(\h^\pe)^2+(\h_\pa)^2, 
\quad
|\HH|^2= \tr(\HHbar\HH) =2|\alpha|^2( |\hh^{2\pe}|^2|\hh^{1\pe}|^2 - |\hhbar{}^{2\pe}\cdot\hh^{1\pe}|^2).
\eeq
Then the conservation law \eqref{so.sg.conslaw} becomes
\beq
|\hh^{1\pe}|^2+|\hh^{2\pe}|^2 +|\alpha|^{-2}
+|\alpha|^2\big(|\hh^{2\pe}|^2|\hh^{1\pe}|^2-|\hhbar{}^{2\pe}\cdot\hh^{1\pe}|^2\big)
=1,
\eeq
which is a quadratic equation for $|\alpha|^{-2}$. 
If we write 
\beq
\beta =|\hh^{2\pe}|^2|\hh^{1\pe}|^2 -|\hhbar{}^{2\pe}\cdot\hh^{1\pe}|^2,
\quad
\gamma =1-|\hh^{1\pe}|^2-|\hh^{2\pe}|^2,
\eeq
then the solution is given by 
\beq
|\alpha|^{-2} =\tfrac{1}{2}(\gamma+\sqrt{\gamma^2-4\beta})
\eeq
with $\beta\geq 0$ due to the Cauchy-Schwartz inequality, 
and $\gamma>0$ due to the conservation law \eqref{so.sg.conslaw}. 
(Note a choice of sign for the square root has been made so that $|\alpha|$ is non-singular for all $\beta\geq 0$.)
This determines 
\beq
\h_\pa=\pm\tfrac{1}{2}\sqrt{2(\gamma+\sqrt{\gamma^2-4\beta})- (\hh^\pe)^2}.
\eeq
As a result, 
the flow equations \eqref{so.sg.ueq}--\eqref{so.sg.u2eq} become a local SG system
with 
\beq
\tilde\h_\pa=\pm\tfrac{1}{2}\sqrt{2(\tilde\gamma+\sqrt{\tilde\gamma^2-4\tilde\beta})- \uu_t^2},
\quad
\tilde\HH_\pa = \tfrac{1}{4}(\i\u_t +2\tilde\h_\pa)\Big(\tilde\gamma+\sqrt{\tilde\gamma^2-4\tilde\beta}\Big) \wedg{\uu_{1 t}}{\uu_{2 t}},
\eeq
where
\beq
\tilde\beta =|\uu_{1 t}|^2|\uu_{2 t}|^2 -|\uubar_{1 t}\cdot\uu_{2 t}|^2,
\quad
\tilde\gamma =\efacsq\lambda^2-|\uu_{1 t}|^2-|\uu_{2 t}|^2.
\eeq

\subsection{Hierarchies of integrable systems}

The mKdV system \eqref{so.mkdv.ueq}--\eqref{so.mkdv.u2eq}
and the NLS system \eqref{so.nls.ueq}--\eqref{so.nls.u2eq}
are each a root system in a hierarchy of integrable systems
generated by the recursion operator \eqref{so.Rop}. 

\begin{thm}\label{thm:SO.hierarchy}
There is a mKdV hierarchy of integrable systems 
\beq
\bpm \u_t\\ \uu_{1t} \\ \uu_{2t} \epm -\lambda\efac \bpm \u_x\\ \uu_{1x} \\ \uu_{2x} \epm 
=\Rop^k \bpm \u_x\\ \uu_{1x} \\ \uu_{2x} \epm 
\quad
k=0,1,2,\ldots
\eeq
as well as a NLS hierarchy of integrable systems 
\beq
\bpm\u_t\\ \uu_{1t} \\ \uu_{2t} \epm -\lambda\efac \bpm 0\\ \i\uu_{1x} \\ \i\uu_{2x} \epm 
=\Rop^k \bpm 0 \\ \i\uu_{1x} \\ \i\uu_{2x} \epm 
\quad
k=0,1,2,\ldots
\eeq
arising from the structure equations \eqref{so.hpar.eq}--\eqref{so.uu2.eq} 
of a $U(n)$-parallel frame for non-stretching curve flows in $SO(2n)/U(n)$. 
Associated to the mKdV hierarchy is an integrable SG system \eqref{so.sg.ueq}--\eqref{so.sg.u2eq}. 
All of these integrable systems are invariant under the unitary symmetry group $U(n-2)\times SU(2)$,
with the $U(n-2)$ subgroup acting as 
$(\u,\uu_1,\uu_2)\to (\u,\uu_1X_{n-2}^{-1},\uu_2X_{n-2}^{-1})$
where $X_{n-2}$ is an arbitrary $x$-independent $(n-2)\times(n-2)$ unitary matrix,
and with the $SU(2)$ subgroup acting as 
$(\uu_1,\uu_2)\to (\uu_1,\uu_2)X_{2}^{-1}$
where $X_{2}$ is an arbitrary $x$-independent $2\times 2$ unimodular unitary matrix. 
\end{thm}

\section{Geometric curve flows}
\label{sec:curveflows}

The hierarchies of integrable systems derived in Theorems~\ref{thm:SU.hierarchy} and~\ref{thm:SO.hierarchy}
correspond to geometrical non-stretching curve flows $\map(x,t)$ 
in the respective symmetric spaces $M=SU(n+1)/U(n)$ and $M=SO(2n)/U(n)$. 
These curve flows can be constructed through the soldering relations \eqref{curveflow}--\eqref{dual.frenet.eq}
for a $U(n)$-parallel frame formulated in Propositions~\ref{prop:SU.results} and~\ref{prop:SO.results}. 

Specifically, 
from the frame structure equations \eqref{cartaneq.mpar}--\eqref{cartaneq.hperp}, 
the flow vector $\map_t$ has the frame components 
$e\hook \map_t = h_\pe+h_\pa$
with $h_{\pa} =D_x^{-1}[h_{\pe},u]_{\pa}$,
where $h_\pe$ is a specified function of $x$, $u$, and $x$-derivatives of $u$,
and where $u$ is Cartan matrix of the $U(n)$-parallel frame along the curve $\map$. 
The curve flow equation is reconstructed from these frame variables by 
the soldering relation
\begin{equation}\label{curveflow.hperp.eq}
\map_t=-K(e^*,\mathcal{Y}(h_{\pe})),
\end{equation}
where $e^*$ is the linear frame dual to the linear coframe $e$, 
$K$ is the Cartan-Killing inner product, 
and $\mathcal{Y}$ is the linear operator
\begin{equation}
\mathcal{Y}:=\id - D_x^{-1}\ad(u)_{\pa} .
\end{equation}
Here $e^*$ is understood to be determined in terms of $u$ 
by the dual Frenet equation \eqref{dual.frenet.eq} along $\map$,
up to the action of the equivalence group $U(n)_\pa$ of the $U(n)$-parallel frame, 
under which $e\hook \map_x=\e$ is preserved. 

We will now derive the curve flow equations corresponding to 
the mKdV system, the NLS system, and the SG system in each symmetric space
$M=SU(n+1)/U(n)$ and $M=SO(2n)/U(n)$. 
These curve flows can be expressed geometrically in a $G$-invariant form 
using the tangent vector $X=\map_x$ 
and the principal normal vector $N=\nabla_x\map_x$ of the curve,
plus the Riemannian metric tensor $g(\cdot,\cdot)$
and the Riemannian curvature tensor $R(\cdot,\cdot)$ on $M=G/U(n)$. 
In general, 
a curve flow equation \eqref{curveflow.hperp.eq} 
in $M=G/U(n)$ will be $G$-invariant 
if and only if $h_{\pe}$ is an equivariant function of 
$x$, $u$, and $x$-derivatives of $u$ 
under the frame equivalence group $U(n)_\pa\subset U(n)$. 

To proceed, 
we will need to use a decomposition of the tangent space $T_\map M$ 
adapted to the tangent vector $\map_x$ of the curve 
in the following geometrical manner. 

We start with the soldering identification $T_x M\simeq \msp$ 
provided by the $U(n)$-parallel frame $e$ along $\map$. 
Recall that the vector space $\msp$ has an orthogonal decomposition 
$\msp=\msp_\pa\oplus \msp_\pe$ 
where $\msp_\pa$ is the centralizer subspace of $\e=e\hook\map_x \in \msp$
and $\msp_\pe$ is the perp subspace. 
We can construct a projection operator onto $\msp_\pe$ in $\msp$
by using the linear map $\ad(\e)^2:\msp\to \msp$ whose null space is $\msp_\pa$. 
When restricted to $\msp_\pe$,
this linear map has two eigenspaces, 
with eigenvalues $-4\efacsq$ and $-\efacsq$,
where 
\beq
\norme =
\begin{cases}
2\sqrt{n+1}, & G= SU(n+1)
\\
4\sqrt{n-1}, &  G=SO(2n)
\end{cases}
\eeq
(see Lemma~\ref{lem:su.struct} in Appendix~\ref{app:su.liealg} 
and Lemma~\ref{lem:so.struct} in Appendix~\ref{app:so.liealg}). 
This determines 
\begin{equation}\label{proj.m.pe}
\mathcal{P}_\pe = (-\tfrac{1}{4}\norme^2)\big( (5\efacsq)\id +\ad(\e)^2 \big)\ad(\e)^2,
\end{equation}
which satisfies $\mathcal{P}_\pe\msp_\pe = \msp_\pe$ and $\mathcal{P}_\pe\msp_\pa =0$. 
A complementary projection operator onto $\msp_\pa$ in $\msp$ is then given by 
\begin{equation}\label{proj.m.pa}
\mathcal{P}_\pa = \id - \mathcal{P}_\pe,
\end{equation}
which satisfies $\mathcal{P}_\pa\msp_\pa = \msp_\pa$ and $\mathcal{P}_\pa\msp_\pe =0$. 

Along the curve $\map$, 
the linear map on $T_x M$ corresponding to $\ad(\e)^2$ on $\msp$ 
is given by 
\begin{equation}\label{adsqmap}
\ad^2(\map_x) = -R(\cdot\,,\map_x)\map_x
\end{equation}
since $e\hook\ad^2(\map_x)Z = -e\hook R(Z,\map_x)\map_x= [[e\hook Z,e\hook\map_x],e\hook\map_x]= \ad(\e)^2(e\hook Z)$ 
holds for all $Z\in T_\map M$ through the soldering relation \eqref{curv.tensor.M}. 
Then we have the projection operators
\begin{align}
\mathcal{P}_\pe & = (-\tfrac{1}{4}\norme^2)\big( (5\efacsq)\id +\ad(\map_x)^2 \big)\ad(\map_x)^2
\label{proj.TM.perp}
\\
\mathcal{P}_\pa & = \id - \mathcal{P}_\pe
\label{proj.TM.par}
\end{align}
where $\norme$ has a geometrical meaning 
coming from the eigenvalues of $R(\cdot\,,\map_x)\map_x$. 
Hence, we have established the following useful result. 

\begin{lem}\label{tangent.sp.decomp} 
For any arclength-parameterized curve $\map$ in $M=G/U(n)$ for
$G=SU(n+1),SO(2n)$, 
the projection operators \eqref{proj.TM.perp}--\eqref{proj.TM.par}
yield a geometrical decomposition 
\beq\label{TM.pe.pa}
T_\map M = (T_\map M)_\pe \oplus (T_\map M)_\pa
\eeq
with 
\beq
\mathcal{P}_\pe T_\map M = (T_\map M)_\pe,
\quad
\mathcal{P}_\pa T_\map M = (T_\map M)_\pa ,
\quad
g((T_\map M)_\pe,(T_\map M)_\pa)=0 
\eeq
which depends only the tangent vector $\map_x$ of the curve
and the Riemannian metric and curvature of the symmetric space $G/U(n)$. 
\end{lem}

This decomposition \eqref{TM.pe.pa} is not preserved by 
the Hermitian structure of $M=G/U(n)$, 
which is given by $J= \Ad(U(1)_C)$ where $U(1)_C$ is the center of $U(n)$. 
However, 
the center of the frame equivalence group $U(n)_\pa\subset U(n)$
is a circle group $U(1)_\pa$ whose action $Ad(U(1)_\pa)$ on $T_\map M$ 
commutes with $\ad(\map_x)^2$. 
This linear map $Ad(U(1)_\pa)$ defines an almost complex structure 
$j_{\map_x}$ in $T_\map M$. 
In the $U(n)$-parallel frame, it is given by 
\begin{equation}\label{J.map}
e\hook j_{\map_x}(Z) = -\ad(\j)(e\hook Z)
\end{equation}
for all $Z\in T_\map M$. 
On the eigenspaces of $\ad(\map_x)^2$ in $(T_\map M)_\pe$, 
with eigenvalues $-4\efacsq$ and $-\efacsq$,
the linear map $j_{\map_x}^2$ has eigenvalues $0,-1$. 
In particular, on the $-1$ eigenspace, 
$j_{\map_x}$ coincides with $J$. 
(On $(T_\map M)_\pa$, which is the null space of  $\ad(\map_x)^2$, 
$j_{\map_x}^2((T_\map M)_\pa) =0$.)

We now apply Lemma~\ref{tangent.sp.decomp} 
to the curve flow equation \eqref{curveflow.hperp.eq}, 
yielding
\beq
\map_t = (\map_t)_\pe + (\map_t)_\pa,
\quad
(\map_t)_\pe = -K(e^*,h_{\pe}),
\quad
(\map_t)_\pa = -K(e^*,\mathcal{Y}(h_{\pe})_\pa),
\eeq
which shows that the perp component $(\map_t)_\pe$ 
determines the entire flow vector $\map_t$. 
This relationship is a consequence of the non-stretching property of the flow 
combined with the properties of a $U(n)$-parallel frame. 
Therefore, hereafter, we will only look at the perp component $(\map_t)_\pe$. 

As a final preliminary step, 
we recall that 
\beq\label{solder.rels}
e\hook \map_x = \e\in \msp_\pa,
\quad
e\hook \nabla_x\map_x = [u,\e] = -\ad(\e)u \in \msp_\pe
\eeq
from the Frenet equation \eqref{frenet.eq} and the Cartan matrix equation \eqref{pull.tangential}.

\subsection{NLS curve flows}

In both symmetric spaces, 
the NLS-type system \eqref{ad_j.flow} is given by $\ad(\e)h_\pe=\h^\pe = \ad(j)u$,
where $\ad(j)$ corresponds to the almost complex structure $j_{\map_x}$ 
through the soldering identification \eqref{J.map}. 
This determines
\beq
h_\pe=\ad(\e)^{-1}\ad(j)u = \ad(\e)^{-2}\ad(j)\ad(\e)u
\eeq
since $\ad(\e)$ and $\ad(\j)$ commute. 
Hence we obtain 
\beq
(e\hook \map_t)_\pe=e\hook( -\ad(\map_x)^{-2} j_{\map_x}(\nabla_x\map_x) ),
\eeq
which gives the NLS curve flow equation 
\beq\label{nls.curveflow}
(\map_t)_\pe=-\ad(\map_x)^{-2} j_{\map_x}(\nabla_x\map_x).
\eeq
This is a geometrical non-stretching curve flow 
in $M=SU(n+1)/U(n)$ and $M=SO(2n)/U(n)$. 
It is a generalization of the bi-normal flow equation
$\map_t = \kappa \hat B$ in Euclidean space $\Rnum^3$, 
since if bi-normal vector is expressed as $\hat B = \hat T\times \hat N$
then the flow equation becomes $\map_t = \map_x\times N$
where $N=\kappa\hat N =\map_{xx}$ is the principal normal vector 
and $\map_x=\hat T$ is the tangent vector. 
In this formulation of the bi-normal equation,
we see that the linear map $\map_x\times$ can be viewed as 
an almost complex structure since $(\map_x\times)^2= -\id$ 
by the standard vector cross product identity,
and hence is a counterpart of $j_{\map_x}$ in the case of Euclidean space.

\subsection{mKdV curve flows}

In both symmetric spaces, 
the mKdV-type system \eqref{x.flow} is given by $\ad(\e)h_\pe=\h^\pe = u_x$. 
This determines
\beq
h_\pe=\ad(\e)^{-1}u_x = \ad(\e)^{-2}\ad(\e)u_x.
\eeq
We can relate $\ad(\e)u_x$ to the derivative of the principal normal vector $N=\nabla_x\map_x$ through the soldering relation \eqref{conn.M},
which gives
\beq
e\hook \nabla_x N = \ad(\e)u_x +[u,\ad(\e)u].
\eeq
The commutator term $[u,\ad(\e)u]$ can be simplified 
by using the Lie brackets \eqref{liebrac3} and \eqref{liebrac4} 
in the symmetric spaces $M=SU(n+1)/U(n)$ and $M=SO(2n)/U(n)$
(see Lemma~\ref{lem:su.struct} in Appendix~\ref{app:su.liealg} 
and Lemma~\ref{lem:so.struct} in Appendix~\ref{app:so.liealg}). 
This yields
\beq
[u,\ad(\e)u] = -K(\ad(\e)u,\ad(\e)u)\e= g(N,N) e\hook\map_x
\eeq
for both symmetric spaces. 
Hence, we have
\beq
\ad(\e)u_x = e\hook( \nabla_x N -g(N,N)\map_x ) 
\eeq
and consequently 
\beq
e\hook (\map_t)_\pe=e\hook( \ad(\map_x)^{-2}(\nabla_x N -g(N,N)\map_x) ).
\eeq
Thus we obtain the mKdV curve flow equation 
\beq\label{mkdv.curveflow}
(\map_t)_\pe=\ad(\map_x)^{-2}(\nabla_x\map_x -g(\map_x,\map_x)\map_x) .
\eeq
This is a geometrical non-stretching curve flow 
in $M=SU(n+1)/U(n)$ and $M=SO(2n)/U(n)$. 
p

\subsection{SG curve flows}

The SG-type system \eqref{kerJ.flow} in both symmetric spaces
is given by $\varpi^\pe=0$, 
which implies $\varpi^\pa=0$ from the frame structure equation \eqref{cartaneq.hpar}. 
Hence we have $\varpi=0$,
and so the connection matrix in the flow direction vanishes, 
\beq
\conx\hook\map_t=0 . 
\eeq
This can be expressed geometrically by observing 
\beq
0=\ad(\e)\conx\hook\map_t=-[\conx\hook\map_t,\e]=e\hook \nabla_t\map_x
\eeq
through the soldering relation \eqref{conn.M}, since $D_t\e=0$. 
As a result, we obtain the SG curve flow equation 
\begin{equation}\label{sg.curveflow}
\nabla_t\map_x =0 . 
\end{equation}
This can be recognized as being a non-stretching wave map equation  
in $M=SU(n+1)/U(n)$ and $M=SO(2n)/U(n)$. 

In the two symmetric spaces, 
the SG system possesses the respective conservation laws \eqref{su.sg.conslaw} and \eqref{so.sg.conslaw}. 
These conservation laws take the form $D_x K(h,h) =0$. 
Using $K(h,h)=-g(\map_t,\map_t)$, we obtain the corresponding conservation law
\beq
\nabla_x|\gamma_t|=0
\eeq
for the SG curve flow. 
Thus, up to a conformal scaling of $t$, 
the SG curve flow equation \eqref{sg.curveflow} describes a flow with unit speed, $|\gamma_t|=1$.

\section{Concluding remarks}
\label{sec:remarks}

The present paper completes a series of work in which 
the general theory developed in \Ref{Anc2008} 
for parallel frames, Hasimoto variables, 
and integrable systems of mKdV type as well as Sine-Gordon (SG) type 
arising from geometrical non-stretching curve flows 
has been applied to all of the simplest types of Riemannian symmetric spaces. 

One new development that we have introduced here is an extension the theory to 
obtain integrable systems of NLS-type 
by exploiting a $U(1)$ subgroup given by the center of the unitary equivalence group of 
the $U(n)$-parallel frame in the symmetric spaces 
$M=SU(n+1)/U(n)$ and $M=SO(2n)/U(n)$. 
In the case of $M=SU(n+1)/U(n)$, 
this leads to a scalar-vector version of the Yajima-Oikawa system \cite{YajOik,Tsu},
whereas in the case of $M=SO(2n)/U(n)$, 
we obtain a novel nonlocal NLS system. 

For future work, we plan to extend this development as far as possible 
to general Riemannian symmetric spaces.

\appendix
\section{Notation}
\label{app:notation}

For vectors $\aa,\bb\in\Cnum^n$,
we denote the standard vector dot product as $\aa\cdot\bb=\aa\bb^\t$, 
and the Hermitian inner product as $\brack{\aa,\bb}=\re(\aabar\cdot\bb)$,
where ``$\t$'' denotes the transpose. 
The tensor product of two vectors is given by $\aa\otimes\bb = \aa^t\bb$,
which is related to the dot product by $\tr(\aa\otimes\bb) = \aa\cdot\bb$,
where ``$\tr$'' denotes the trace. 

We define the following useful outer products:
\begin{align}
&\wedg{\aa}{\bb}=\aa^\t\bb-\bb^\t\aa
\in\mk{so}(n,\Cnum) .
\label{wedgprod}
\\
&\symm{\aa}{\bb}=\aa^\t\bb+\bb^\t\aa
\in \mk{s}(n,\Cnum),
\label{symmprod}
\\
&\wedgbar{\aa}{\bb}=\aabar^\t\bb-\bbbar^\t\aa
\in\mk{u}(n),
\label{wedgbarprod}
\\
&\symmbar{\aa}{\bb}=\aabar^\t\bb+\bbbar^\t\aa
\in \i\mk{u}(n),
\label{symmbarprod}
\end{align}
where $\i\mk{u}(n)$ denotes the vector space of $n\times n$ hermitian matrices,
and $\mk{s}(n)$ denotes the vector space of $n\times n$ symmetric matrices. 
These outer products have the properties 
\begin{gather}
\wedg{\aa}{\bb} = -\wedg{\bb}{\aa} ,
\quad
(\wedg{\aa}{\bb})^\t = \wedg{\bb}{\aa} ,
\\
\symm{\aa}{\bb} = \symm{\bb}{\aa} ,
\quad
(\symm{\aa}{\bb})^\t = \symm{\bb}{\aa} ,
\\
\wedgbar{\aa}{\bb} = -\wedgbar{\bb}{\aa} ,
\quad 
(\wedgbar{\aa}{\bb})^\t = \wedgbar{\bbbar}{\aabar} ,
\\
\symmbar{\aa}{\bb} = \symmbar{\bb}{\aa} ,
\quad
(\symmbar{\aa}{\bb})^\t = \symmbar{\bbbar}{\aabar} 
\end{gather}
and
\begin{gather}
\tr(\symmbar{\aa}{\bb})=2\brack{\aa,\bb},
\\
\i\tr(\wedgbar{\aa}{\bb})=2\im(\bbbar\cdot\aa)= -2\im(\aabar\cdot\bb)
%\tr(\symmbar{\aa}{\bb}) + \tr(\wedgbar{\aa}{\bb})= 2\aabar\cdot \bb.
\end{gather}

We likewise define the contraction between a vector and a matrix by 
\beq
\cc\hook(\aa\otimes\bb) = (\aa\cdot\cc)\bb,
\quad
(\aa\otimes\bb)\hook\cc = (\bb\cdot\cc)\aa,
\eeq
and hence
\begin{align}
& \cc\hook(\wedg{\aa}{\bb}) = (\cc\cdot\aa)\bb - (\cc\cdot\bb)\aa,
\quad
\cc\hook(\symm{\aa}{\bb}) = (\cc\cdot\aa)\bb + (\cc\cdot\bb)\aa,
\\
& \cc\hook(\wedgbar{\aa}{\bb}) = (\cc\cdot\aabar)\bb - (\cc\cdot\bbbar)\aa,
\quad
\cc\hook(\symmbar{\aa}{\bb}) = (\cc\cdot\aabar)\bb + (\cc\cdot\bbbar)\aa.
\end{align}

We also define the full contraction between two matrices by 
\beq
\AA\cdot\BB = \tr(\AA\BB) . 
\eeq
Finally, we extend the Hermitian inner product to matrices by defining 
\beq
|\AA|^2 = \AAbar\cdot\AA = \tr(\AAbar\AA) . 
\eeq

Some useful identities:
\begin{gather}
(\aa \otimes \bb) \cdot \AA
= \aa \cdot (\AA \hook \bb) 
= \bb \cdot (\aa \hook \AA) ,
\\
\BB\hook\aa = -\aa\hook\BB,
\\
\CC\hook\aa = \aa\hook\CC , 
\end{gather}
for $\AA\in\mk{gl}(n,\Cnum)$, $\BB\in\mk{so}(n,\Cnum)$, $\CC\in\mk{s}(n,\Cnum)$.

\section{Symmetric Lie algebra $\mk{su}(n+1)/\hsp$}
\label{app:su.liealg}

In the matrix representation \eqref{su.matr.rep}--\eqref{su.msp.matr.rep} 
of the Lie algebra $\mk{su}(n+1)=\hsp\oplus\msp$, 
the Lie brackets are given by 
\begin{align}
[\hsp,\msp] & =[\lrep\BB_1\rrep,\lrep\aa_2\rrep]=\lrep\aa_3\rrep\in\msp, 
\quad
\aa_3 = -\tr(\BB_1)\aa_2-\aa_2\BB_1,
\\
[\msp,\msp] &=[\lrep\aa_1\rrep,\lrep\aa_2\rrep]=\lrep\BB_3\rrep\in\hsp, 
\quad
\BB_3= \wedgbar{\aa_2}{\aa_1},
\\
[\hsp,\hsp] &=[\lrep\BB_1\rrep,\lrep\BB_2\rrep]=\lrep[\BB_1,\BB_2]\rrep\in\hsp.
\end{align}

The vector space $\msp=\mk{su}(n+1)/\mk{u}(n) \simeq \Cnum^n$ 
has the following properties.
\begin{prop}\label{prop:su.struct}
\hfil \newline
1. The restriction of the Cartan-Killing form on $\mk{su}(n+1)$ to $\msp$
yields a negative-definite inner product
\beq
K(\lrep\aa\rrep,\lrep\bb\rrep) = 2(n+1)\tr\left(\bpm 0&\aa\\-\aabar^\t & \mb{0} \epm \bpm 0&\bb\\-\bbbar^\t & \mb{0} \epm \right) 
%= -2(n+1)\tr(\symmbar{\aa}{\bb})
=-4(n+1)\brack{\aa,\bb}.
\eeq
\hfil \newline
2. The rank of $\msp$ is $1$,
and there is a Hermitian structure 
$\JJ:= -\tfrac{1}{n+1}\lrep \i I_n\rrep \in\hsp$ 
%\tfrac{1}{n+1}\bpm n\i & \mb{0} \\0& -\i I_n \epm 
which acts by 
$\ad_\JJ\lrep\aa\rrep=[\JJ,\lrep\aa\rrep] = \lrep\i\aa\rrep$.
\hfil \newline
3. Up to isomorphism, 
the Cartan subspace $\asp\subset\msp$ is given by the real span of 
any vector $\e\in\msp$ such that $-K(\e,\e) =1$, 
with $\asp=\spn(\e)_\Rnum$. 
This vector can be chosen as 
\beq
\norme\e= \lrep\ee\rrep,
\quad
\ee = (1,\mb{0})\in\Cnum^n = \Cnum\oplus\Cnum^{n-1}\simeq\msp,
\quad 
\mb{0}\in\Cnum^{n-1},
\eeq
where $\norme^2= -K(\lrep\ee\rrep,\lrep\ee\rrep)= 4(n+1) \brack{\ee,\ee} = 4(n+1)$ 
determines the normalization constant 
\beq\label{su.e.norm}
\norme = 2\sqrt{n+1} .
\eeq
\hfil\newline
4. The Cartan subspace is not invariant under the Hermitian structure, 
since $\ad_\JJ(\e) = \i\e \not\in\spn(\e)_\Rnum$. 
\end{prop}

The Cartan element $\lrep\ee\rrep$ produces 
a decomposition of $\msp=\mpasp\oplus\mpesp$ and $\hsp=\hpasp\oplus\hpesp$ 
into respective centralizer subspaces and perp subspaces
defined by the matrix representations 
\eqref{su.msp.perp.pa.matr.rep} and \eqref{su.hsp.perp.pa.matr.rep}. 
These subspaces have the following main properties. 

\begin{lem}\label{lem:su.struct}
\hfil \newline
1. $\hsp_\pa\simeq \mk{u}(n-1)$ is the centralizer subalgebra of $\e$, 
and $\hsp_\pe \simeq \i\Rnum\oplus\Cnum^{n-1}$ is its perp space; 
$\msp_\pa\simeq \Rnum$ is the centralizer subspace of $\e$, 
and $\msp_\pe \simeq \i\Rnum\oplus\Cnum^{n-1}$ is its perp space. 
\hfil \newline
2. The Lie bracket $[\msp,\msp]$ has the decomposition 
\begin{align}
&
[\lrep\a_\pa\rrep,\lrep\c_\pa\rrep]
=0
\in\h_\pa,
\\
& 
[\lrep\a_\pa\rrep,\lrep(\i \a_\pe,\aa_\pe)\rrep]
=\lrep(-2\i\a_\pa\a_\pe,\,-\a_\pa\aa_\pe)\rrep
\in\h_\pe,
\\
&
[\lrep(\i\a_\pe,\aa_\pe),\lrep(\i\c_\pe,\cc_\pe)]_\pa
=
\lrep 
\wedgbar{\cc_\pe}{\aa_\pe} 
\rrep
\in\h_\pa,
\\
&
[\lrep(\i \a_\pe,\aa_\pe)\rrep,\lrep(\i \c_\pe,\cc_\pe)\rrep]_\pe
=
\lrep 
(\tfrac{1}{2} \tr(\wedgbar{\cc_\pe}{\aa_\pe}),\i\a_\pe\cc_\pe-\i\c_\pe\aa_\pe)
\rrep
\in\h_\pe.
\end{align}
3. The Lie bracket $[\msp,\hsp]$ has the decomposition 
\begin{align}
&
[\lrep\a_\pa\rrep,\lrep\BB_\pa\rrep]
=0
\in\h_\pa,
\\
&
[\lrep\a_\pa\rrep,\lrep(\i\b_\pe,\bb_\pe)\rrep]
= \lrep(2\i\a_\pa\b_\pe,\a_\pa\bb_\pe)\rrep
\in\msp_\pe,
\\
&
[\lrep(\i \a_\pe,\aa_\pe)\rrep,\lrep\BB_\pa\rrep]
=\lrep(0,\tfrac{1}{2}\tr(\BB_\pa)\aa_\pe+\aa_\pe\BB_\pa)\rrep
\in\msp_\pe,
\\
&
[(\i \a_\pe,\aa_\pe),(\i \b_\pe,\bb_\pe)]_\pa
=
\lrep
(-2\a_\pe\b_\pe -\brack{\aa_\pe,\bb_\pe})
\rrep
\in\msp_\pa,
\\
&
[(\i \a_\pe,\aa_\pe),(\i \b_\pe,\bb_\pe)]_\pe
=
\lrep
(-\tfrac{1}{2} \tr(\wedgbar{\bb_\pe}{\aa_\pe}),\i\b_\pe\aa_\pe+\i\a_\pe\bb_\pe)
\rrep
\in\msp_\pe.
\end{align}
4. The Lie bracket $[\hsp,\hsp]$ has the decomposition 
\begin{align}
&
[\lrep\BB_\pa\rrep,\lrep\DD_\pa\rrep]
=\lrep[\BB_\pa,\DD_\pa]\rrep
\in\h_\pa,
\\
&
[\lrep\BB_\pa\rrep,\lrep(\i \b_\pe,\bb_\pe)\rrep]
=\lrep(0,-\tfrac{1}{2}\tr(\BB_\pa)\bb_\pe-\bb_\pe\BB_\pa)\rrep
\in\h_\pe,
\\
&
[(\i \b_\pe,\bb_\pe),(\i \d_\pe,\dd_\pe)]_\pa
=\lrep
\wedgbar{\dd_\pe}{\bb_\pe}
\rrep
\in\h_\pa,
\\
&
[(\i \b_\pe,\bb_\pe),(\i \d_\pe,\dd_\pe)]_\pe
=\lrep
(-\tfrac{1}{2} \tr(\wedgbar{\dd_\pe}{\bb_\pe}),\i\b_\pe\dd_\pe-\i\d_\pe\bb_\pe)
\rrep
\in\h_\pe.
\end{align}
5. The center of the centralizer subalgebra $\hsp_\pa\simeq \mk{u}(n-1)$ of $\e$
is a $\mk{u}(1)$ subalgebra generated by 
$\j:=\tfrac{2}{2n-1}\lrep \i I_{n-1}\rrep \in \hsp_\pa$ 
which acts on $\msp_\pe$ and $\msp_\pa$ by 
\beq
\ad(\j)\lrep(\i\a_\pe,\aa_\pe)\rrep 
= \lrep(0,\i\aa_\pe)\rrep ,
\quad
\ad(\j)\lrep\a_\pa\rrep =0.
\eeq
6. There are vector-space isomorphisms 
\begin{align}
& 
\msp_\pe \mapsto \hsp_\pe: 
\ad(\e)\lrep(\i\a_\pe,\aa_\pe)\rrep
=-\efac \lrep(2\i\a_\pe,\aa_\pe)\rrep,
\\
&
\hsp_\pe \mapsto \msp_\pe: 
\ad(\e)\lrep(\i\b_\pe,\bb_\pe)\rrep
=\efac \lrep(2\i\b_\pe,\bb_\pe)\rrep,
\\
& 
\msp_\pe \mapsto \msp_\pe: 
\ad(\e)^2\lrep(\i\a_\pe,\aa_\pe)\rrep
=-\efacsq \lrep(4\i\a_\pe,\aa_\pe)\rrep,
\\
&
\hsp_\pe \mapsto \hsp_\pe: 
\ad(\e)^2\lrep(\i\b_\pe,\bb_\pe)\rrep
=-\efacsq \lrep(4\i\b_\pe,\bb_\pe)\rrep.
\end{align}
7. Under the Hermitian structure, 
$(\ad_\JJ\lrep(\i\a_\pe,\aa_\pe)\rrep)_\pe = \lrep(0,\i\aa_\pe)\rrep$
and $(\ad_\JJ\lrep(\i\a_\pe,\aa_\pe)\rrep)_\pa =  \lrep(-a_\pe,\mb{0})\rrep$.
\end{lem}

The Lie subalgebra $\hpasp$ generates a group 
$U(n)_\pa = U(n-1)\subset U(n)$ that leaves $\e$ invariant in $\msp$, 
$\Ad(U(n)_\pa)\e = \e$. 

\begin{lem}\label{lem:su.equiv.group}
The invariance group $U(n)_\pa = U(n-1)\subset U(n)$ of $\e$ 
is given by the matrix representation
\beq\label{su.equiv.group}
X = 
\bpm 
\det(X_{n-1})^{-1/2}& (0, \mb{0}) \\ 
(0,\mb{0})^\t  & 
\bpm \det(X_{n-1})^{-1/2}& \mb{0} \\ \mb{0}^\t & X_{n-1} \epm 
\epm 
\in SU(n+1), 
\quad
X_{n-1}\in U(n-1) . 
\eeq
This group acts on $\msp_\pe$ by 
\beq\label{su.equiv.group.action}
\Ad(X)\lrep(\i\a_\pe,\aa_\pe) \rrep
=\lrep(\i\a_\pe,\det(X_{n-1})^{-1/2}\aa_\pe X_{n-1}^{-1})\rrep
\in\msp_\pe 
\eeq
and it leaves invariant $\lrep\a_\pa\rrep \in \msp_\pa$. 
\end{lem}

The center of $U(n)_\pa$ is a $U(1)$ subgroup given by the matrix representation 
\beq\label{su.equiv.group.center}
Z = \exp(\phi\ad_\j) 
= \bpm 
e^{-\i\phi/2} & (0, \mb{0}) \\ 
(0,\mb{0})^\t  & 
\bpm e^{-\i\phi/2} & \mb{0} \\ \mb{0}^\t & e^{\i \phi} I_{n-1} \epm 
\epm 
\in SU(n+1),
\quad
\phi\in\Rnum.
\eeq

\section{Symmetric Lie algebra $\mk{so}(2n)/\hsp$}
\label{app:so.liealg}

In the matrix representation \eqref{so.matr.rep}--\eqref{so.msp.matr.rep} 
of the Lie algebra $\mk{so}(2n)=\hsp\oplus\msp$, 
the Lie brackets  are given by 
\begin{align}
[\hsp,\msp] & = [\lrep\BB_1\rrep,\lrep\AA_2\rrep]=\lrep\AA_3\rrep\in\msp, 
\quad
\AA_3 = \BBbar_1\AA_2 -\AA_2\BB_1,
\\
[\msp,\msp] & =[\lrep\AA_1\rrep,\lrep\AA_2\rrep]=\lrep\BB_3\rrep\in\hsp, 
\quad
\BB= \AAbar_1\AA_2 - \AAbar_2\AA_1,
\\
[\hsp,\hsp] & =[\lrep\BB_1\rrep,\lrep\BB_2\rrep]=\lrep[\BB_1,\BB_2]\rrep\in\hsp.
\end{align}

The vector space $\msp=\mk{so}(2n)/\mk{u}(n) \simeq \Cnum^{\frac{1}{2}n(n-1)}$ 
has the following properties.

\begin{prop}\label{prop:so.struct}
\hfil \newline
1. The restriction of the Cartan-Killing form on $\mk{so}(2n)$ to $\msp$
yields a negative-definite inner product
\beq
K(\lrep\AA_1\rrep,\lrep\AA_2\rrep)=-4(n-1)\re(\tr(\AA_1\AAbar_2)).
\eeq
\hfil \newline
2. The rank of $\msp$ is $[n/2]$ (integer part), 
and there is a Hermitian structure 
$\JJ:=-\tfrac{1}{2}\lrep \i I_n\rrep \in\hsp$ 
%-\tfrac{1}{2}\bpm 0&I_n\\-I_n&0\epm  
which acts by $\ad_\JJ\lrep\AA\rrep=\lrep\i\AA\rrep$. 
\hfil \newline
3. Up to isomorphism, 
the Cartan subspace $\asp\subset\msp$ is given by 
the real span of the $[n/2]$ matrices
\beq
\e_k:=\lrep E_{2k-1,2k}-E_{2k,2k-1} \rrep\in\msp,
\quad k=1,\ldots ,[n/2],
\eeq
where $E_{i,j}\in\mk{gl}(n,\Cnum)$ denotes the matrix such that its $(i,j)$ entry is $1$ and all other entries are $0$. 
\hfil \newline
4. The Cartan subspace is not invariant under $\JJ$,
due to $\ad_\JJ(\e_k) = \i\e_k \not\in\spn(\e_1,\ldots,\e_{[n/2]})_\Rnum$.
\end{prop}

The basis matrices $\e_k$ have the distinguishing property that the centralizer subspace of each one is of maximal dimension. 
We will now select a unit-norm element $\e$ in the Cartan subspace 
by choosing any one of the basis matrices
\beq
\norme \e = \e_1\in \asp,
\eeq
where 
\beq
-\norme^2= K(\e_1,\e_1) = 8(n-1)\re\tr((E_{1,2}-E_{2,1})^2)= -16(n-1)
\eeq
determines the normalization constant 
\beq\label{so.e.norm}
\norme=4\sqrt{n-1}.
\eeq
This choice of a Cartan element produces 
a decomposition of $\msp=\mpasp\oplus\mpesp$ and $\hsp=\hpasp\oplus\hpesp$ 
into respective centralizer subspaces and perp subspaces
defined by the matrix representations 
\eqref{so.msp.perp.pa.matr.rep} and \eqref{so.hsp.perp.pa.matr.rep}. 
These subspaces have the following main properties. 

\begin{lem}\label{lem:so.struct}
\hfil \newline
1. $\hsp_\pa\simeq \mk{su}(2)\oplus\mk{u}(n-2)$ is the centralizer subalgebra of $\e$, 
and $\hsp_\pe\simeq \i\Rnum\oplus\Cnum^{n-2}\oplus \Cnum^{n-2}$ is its perp space; 
$\msp_\pa\simeq \Rnum\oplus\mk{so}(n-2,\Cnum)$ is the centralizer subspace of $\e$, 
and $\msp_\pe\simeq \i\Rnum\oplus\Cnum^{n-2}\oplus\Cnum^{n-2}$ is its perp space. 
\hfil \newline
2. The Lie bracket $[\msp,\msp]$ has the decomposition 
\begin{align}
& [\lrep(\a_\pa,\AA_\pa)\rrep,\lrep(\c_\pa,\CC_\pa)\rrep]
=\lrep
(0,\mb{0},\AAbar_\pa\CC_\pa-\CCbar_\pa\AA_\pa)
\rrep
\in\hsp_\pa,
\\
& 
[\lrep(\a_\pa,\AA_\pa)\rrep,\lrep(\i\a_\pe,\aaone_\pe,\aatwo_\pe)\rrep]
=\lrep
(-2\i\a_\pa\a_\pe,\a_\pa\aatwo_\pe-\aaonebar_\pe\AA_\pa,
-\a_\pa\aaone_\pe-\aatwobar_\pe\AA_\pa)
\rrep
\in\hsp_\pe,
\\
&\begin{aligned}
& \lbrack\lrep(\i\a_\pe,\aaone_\pe,\aatwo_\pe)\rrep,\lrep(\i\c_\pe,\ccone_\pe,\cctwo_\pe)\rrep\rbrack_\pa \\&
= \lrep(
\tfrac{1}{2}(\tr(\wedgbar{\ccone_\pe}{\aaone_\pe})-\tr(\wedgbar{\cctwo_\pe}{\aatwo_\pe})),
\cconebar_\pe\cdot\aatwo_\pe-\aaonebar_\pe\cdot\cctwo_\pe
,
\wedgbar{\ccone_\pe}{\aaone_\pe}+\wedgbar{\cctwo_\pe}{\aatwo_\pe} 
)\rrep
\in\hsp_\pa,
\end{aligned}
\\
&\begin{aligned}
& \lbrack\lrep(\i\a_\pe,\aaone_\pe,\aatwo_\pe)\rrep,\lrep(\i\c_\pe,\ccone_\pe,\cctwo_\pe)\rrep\rbrack_\pe
\\&
=\lrep( 
\tfrac{1}{2}(\tr(\wedgbar{\ccone_\pe}{\aaone_\pe}+\tr(\wedgbar{\cctwo_\pe}{\aatwo_\pe})),
\i(\c_\pe\aatwo_\pe-\a_\pe\cctwo_\pe),\i(\a_\pe\ccone_\pe-\c_\pe\aaone_\pe) 
)\rrep
\in\hsp_\pe.
\end{aligned}
\end{align}
3. The Lie bracket of $[\msp,\hsp]$ has the decomposition 
\begin{align}
& [\lrep\a_\pa,\AA_\pa\rrep,\lrep\i\bone_\pa,\btwo_\pa,\BB_\pa\rrep]
=\lrep
(0,\AA_\pa\BB_\pa-\BBbar_\pa\AA_\pa)
\rrep
\in\msp_\pa,
\\
&
[\lrep(\a_\pa,\AA_\pa)\rrep,\lrep(\i\b_\pe,\bbone_\pe,\bbtwo_\pe)\rrep]
=\lrep
(2\i\a_\pa\b_\pe,\a_\pa\bbtwo_\pe-\bbonebar_\pe\AA_\pa,
-\a_\pa\bbone_\pe-\bbtwobar_\pe\AA_\pa)
\rrep
\in\msp_\pe,
\\
& \begin{aligned}
& [\lrep(\i\a_\pe,\aaone_\pe,\aatwo_\pe)\rrep,\lrep(\i\bone_\pa,\btwo_\pa,\BB_\pa)\rrep]
\\&
=\lrep
(0,\aaone_\pe\BB_\pa+\i\bone_\pa\aaone_\pe-\btwobar_\pa\aatwo_\pe,
\aatwo_\pe\BB_\pa-\i\bone_\pa\aatwo_\pe+\btwo_\pa\aaone_\pe)
\rrep
\in\msp_\pe,
\end{aligned}
\\
&\begin{aligned}
& \lbrack\lrep(\i\a_\pe,\aaone_\pe,\aatwo_\pe)\rrep,\lrep(\i\b_\pe,\bbone_\pe,\bbtwo_\pe)\rrep\rbrack_\pa
\\& 
=\lrep(
-2\a_\pe\b_\pe+\tfrac{1}{2}(\tr(\symmbar{\bbone_\pe}{\aatwo_\pe})-\tr(\symmbar{\bbtwo_\pe}{\aaone_\pe})),
\wedg{\bbone_\pe}{\aaone_\pe}+\wedg{\bbtwo_\pe}{\aatwo_\pe}
\rrep)
\in\msp_\pa,
\end{aligned}
\\
&\begin{aligned}
& \lbrack\lrep(\i\a_\pe,\aaone_\pe,\aatwo_\pe)\rrep,\lrep(\i\b_\pe,\bbone_\pe,\bbtwo_\pe)\rrep\rbrack_\pe
\\& 
=\lrep(
\tfrac{1}{2}(\tr(\wedgbar{\bbone_\pe}{\aatwo_\pe})-\tr(\wedgbar{\bbtwo_\pe}{\aaone_\pe})),
\i(\a_\pe\bbtwo_\pe+\b_\pe\aaone_\pe),\i(-\a_\pe\bbone_\pe+\b_\pe\aatwo_\pe)
\rrep)
\in\msp_\pe.
\end{aligned}
\end{align}
4. The Lie bracket $[\hsp,\hsp]$ has the decomposition 
\begin{align}
&\begin{aligned}
& [\lrep(\i\bone_\pa,\btwo_\pa,\BB_\pa)\rrep,\lrep(\i\done_\pa,\dtwo_\pa,\DD_\pa)\rrep]
\\&
=\lrep
(\dtwo_\pa\btwobar_\pa -\btwo_\pa\dtwobar_\pa,2\i(\bone_\pa\dtwo_\pa-\done_\pa\btwo_\pa),\BB_\pa\DD_\pa-\DD_\pa\BB_\pa)
\rrep
\in\hsp_\pa,
\end{aligned}
\\
&\begin{aligned}
& [\lrep(\i\bone_\pa,\btwo_\pa,\BB_\pa)\rrep,\lrep(\i\b_\pe,\bbone_\pe,\bbtwo_\pe)\rrep]
\\&
=\lrep
(0,-\bbone_\pe\BB_\pa+\i\bone_\pa\bbone_\pe+\btwo_\pa\bbtwo_\pe,
-\bbtwo_\pe\BB_\pa -\i\bone_\pa\bbtwo_\pe-\btwobar_\pa\bbone_\pe)
\rrep
\in\hsp_\pe,
\end{aligned}
\\
&\begin{aligned}
& \lbrack\lrep(\i\b_\pe,\bbone_\pe,\bbtwo_\pe)\rrep,\lrep(\i\d_\pe,\ddone_\pe,\ddtwo_\pe)\rrep\rbrack_\pa
\\&
=\lrep( 
\tfrac{1}{2}(\tr(\wedgbar{\bbone_\pe}{\ddone_\pe}) -\tr(\wedgbar{\bbtwo_\pe}{\ddtwo_\pe})),
\bbtwobar_\pe\cdot\ddone_\pe-\ddtwobar_\pe\cdot\bbone_\pe,
\wedgbar{\ddone_\pe}{\bbone_\pe}+\wedgbar{\ddtwo_\pe}{\bbtwo_\pe} 
)\rrep
\in\hsp_\pa,
\end{aligned}
\\
&\begin{aligned}
& \lbrack\lrep(\i\b_\pe,\bbone_\pe,\bbtwo_\pe)\rrep,\lrep(\i\d_\pe,\ddone_\pe,\ddtwo_\pe)\rrep\rbrack_\pe
\\&
=\lrep( 
\tfrac{1}{2}(\tr(\wedgbar{\bbone_\pe}{\ddone_\pe})+\tr(\wedgbar{\bbtwo_\pe}{\ddtwo_\pe})),
\i(\b_\pe\ddone_\pe-\d_\pe\bbone_\pe),\i(\b_\pe\ddtwo_\pe-\d_\pe\bbtwo_\pe) 
)\rrep
\in\hsp_\pe.
\end{aligned}
\end{align}
5. The center of the centralizer subalgebra $\hsp_\pa\simeq \mk{su}(2)\oplus\mk{u}(n-2)$ of $\e$
is a $\mk{u}(1)$ subalgebra generated by $\j:=-\lrep\i I_{n-2}\rrep \in \hsp_\pa$ 
which acts on $\msp_\pe$ by 
\beq
\ad(\j)\lrep(\i\a_\pe,\aaone_\pe,\aatwo_\pe)\rrep 
= \lrep(0,\i\aaone_\pe,\i\aatwo_\pe)\rrep .
\eeq
6. There are isomorphisms of vector spaces 
\begin{align}
& 
\msp_\pe \mapsto \hsp_\pe: 
\ad(\e)\lrep(\i\a_\pe,\aaone_\pe,\aatwo_\pe)\rrep
=\efac \lrep(-2\i\a_\pe,\aatwo_\pe,-\aaone_\pe)\rrep,
\\
&
\hsp_\pe \mapsto \msp_\pe: 
\ad(\e)\lrep(\i\b_\pe,\bbone_\pe,\bbtwo_\pe)\rrep
=\efac \lrep(2\i\b_\pe,\bbtwo_\pe,-\bbone_\pe)\rrep.
\end{align}
7. Under the Hermitian structure, 
$(\ad_\JJ\lrep(\i\a_\pe,\aaone_\pe,\aatwo_\pe)\rrep)_\pe = \lrep(0,\i\aaone_\pe,\i\aatwo_\pe)\rrep$
and\\$(\ad_\JJ\lrep(\i\a_\pe,\aaone_\pe,\aatwo_\pe)\rrep)_\pa =  \lrep(-a_\pe,\mb{0})\rrep$.
\end{lem}

The Lie subalgebra $\hpasp$ generates a group 
$U(n)_\pa = SU(2)\times U(n-2)\subset U(n)$ that leaves $\e$ invariant in $\msp$, 
$\Ad(U(n)_\pa)\e = \e$. 

\begin{lem}\label{lem:so.equiv.group}
The subgroups $SU(2)\subset U(n)_\pa$ and $U(n-2)\subset U(n)_\pa$
in the invariance group $U(n)_\pa$ of $\e$ 
are given by the respective matrix representations
\begin{align}
& X_{SU(2)} = 
\bpm 
e^{\i\phi}\cos\theta &e^{\i\phi}\sin\theta& \mb{0}\\
-e^{-\i\phi}\sin\theta& e^{-\i\phi}\cos\theta & \mb{0}\\
\mb{0}^\t&\mb{0}^\t& \mb{0}
\epm 
\simeq SU(2) \in U(n), 
\quad
\phi,\theta\in\Rnum,
\label{so.equiv.SU.group}
\\
& X_{U(n-2)} = 
\bpm 
0 & 0 & \mb{0}\\
0 & 0 &\mb{0}\\
\mb{0}^\t&\mb{0}^\t& X_{n-2}
\epm
\simeq U(n-2) \in U(n), 
\quad
X_{n-2}\in U(n-2) . 
\label{so.equiv.U.group}
\end{align}
These subgroups act on $\msp_\pe$ by 
\begin{align}
& \Ad(X_{SU(2)})\lrep(\i\a_\pe,\aaone_\pe,\aatwo_\pe)\rrep
= \lrep (\i\a_\pe,\mb{0},\mb{0})\rrep + \lrep X_{SU(2)}\rrep \lrep(0,\aaone_\pe,\aatwo_\pe)\rrep
\in\msp_\pe ,
\quad
\label{so.equiv.SU.group.action}
\\
& \Ad(X_{U(n-2)})\lrep(\i\a_\pe,\aaone_\pe,\aatwo_\pe)\rrep
= \lrep(\i\a_\pe,\aaone_\pe X_{n-2}^{-1},\aatwo_\pe X_{n-2}^{-1})\rrep
\in\msp_\pe ,
\end{align}
and they leave invariant $\lrep(\a_\pa,\mb{0})\rrep \in \msp_\pa$. 
Composition of these two subgroups yields the group $U(n)_\pa=SU(2)\times U(n-2) \subset U(n)$. 
\end{lem}

The center of $U(n)_\pa$ is a $U(1)$ subgroup given by the matrix representation 
\beq\label{so.equiv.group.center}
Z = \exp(\phi\ad_\j) 
= \bpm 
0 & 0 & \mb{0}\\
0 & 0 &\mb{0}\\
\mb{0}^\t  & \mb{0}^\t  & e^{-\i \phi} I_{n-2} 
\epm 
\in U(n),
\quad
\phi\in\Rnum.
\eeq

\section*{Acknowledgements}
S.C.A. is supported by an NSERC research grant. 
E.A. thanks the Mathematics \& Statistics Department of Brock University 
for support during the period in which this work was completed. 

Takayuki Tsuchida is thanked for helpful remarks.

\end{document}